\documentclass[twocolumn, 10pt, longbibliography, superscriptaddress, prl, aps]{revtex4-1}
\usepackage{amsmath}
\usepackage{amssymb}
\usepackage{graphicx}
\usepackage{xfrac}
\usepackage{upgreek}
\usepackage{color}
\usepackage{soul}

\usepackage{color}
    \definecolor{darkred}{rgb}{0.5, 0, 0}
    \definecolor{darkgreen}{rgb}{0, 0.5, 0}
    \definecolor{darkblue}{rgb}{0.1, 0.1, 0.8}
\usepackage[bookmarks=true,
            bookmarksnumbered=false,
            bookmarksopen=true,
            breaklinks=true,
            pdfborder={0 0 0},
            final=true,
            backref=none,
            colorlinks=true,
            linkcolor=darkblue,
            menucolor=darkblue,
            citecolor=darkblue,
            urlcolor=darkblue]{hyperref}
            
\renewcommand{\mu}{\upmu}
\newcommand{\um}{\mu\textrm{m}}			
\newcommand{\ium}{\mu\textrm{m}^{-1}} 	

\makeindex

\begin{document}

\title{Transition from propagating polariton solitons to a standing wave condensate induced by interactions}

\author{M.~Sich}
\thanks{m.sich@sheffield.ac.uk}
\affiliation{Department of Physics and Astronomy, The University of Sheffield, Sheffield, S3~7RH, United Kingdom}

\author{J.~K.~Chana}
\affiliation{Department of Physics and Astronomy, The University of Sheffield, Sheffield, S3~7RH, United Kingdom}
\affiliation{Base4 Innovation Ltd, Cambridge, CB3~0FA, United Kingdom}

\author{O.~A.~Egorov}
\affiliation{Technische Physik der Universit\"{a}t W\"{u}rzburg, Am Hubland, 97074, W\"{u}rzburg, Germany}

\author{H.~Sigurdsson}
\affiliation{Science Institute, University of Iceland, Dunhagi-3, IS-107 Reykjavik, Iceland}

\author{I.~A.~Shelykh}
\affiliation{Science Institute, University of Iceland, Dunhagi-3, IS-107 Reykjavik, Iceland}
\affiliation{Department of Nanophotonics and Metamaterials, ITMO University, St. Petersburg, 197101, Russia}

\author{D. V. Skryabin}
\affiliation{Department of Physics, University of Bath, Bath, BA2 7AY, United Kingdom}
\affiliation{Department of Nanophotonics and Metamaterials, ITMO University, St. Petersburg, 197101, Russia}

\author{P.~M.~Walker}
\affiliation{Department of Physics and Astronomy, The University of Sheffield, Sheffield, S3~7RH, United Kingdom}

\author{E. Clarke}
\affiliation{EPSRC National Centre for III-V Technologies, The University of Sheffield, Sheffield, S1~4DE, United Kingdom}

\author{B. Royall}
\affiliation{Department of Physics and Astronomy, The University of Sheffield, Sheffield, S3~7RH, United Kingdom}

\author{M.~S.~Skolnick}
\affiliation{Department of Physics and Astronomy, The University of Sheffield, Sheffield, S3~7RH, United Kingdom}
\affiliation{Department of Nanophotonics and Metamaterials, ITMO University, St. Petersburg, 197101, Russia}

\author{D.~N.~Krizhanovskii}
\thanks{d.krizhanovskii@sheffield.ac.uk}
\affiliation{Department of Physics and Astronomy, The University of Sheffield, Sheffield, S3~7RH, United Kingdom}
\affiliation{Department of Nanophotonics and Metamaterials, ITMO University, St. Petersburg, 197101, Russia}

\begin{abstract}
We explore nonlinear transitions of polariton wavepackets, first, to a soliton and then to a standing wave polariton condensate in a multi-mode microwire system. At low polariton density we observe ballistic propagation of the multi-mode polariton wavepackets arising from the interference between different transverse modes. With increasing polariton density, the wavepackets transform into single mode bright solitons due to effects of both inter-modal and intra-modal polariton-polariton scattering. Further increase of the excitation density increases thermalisation speed leading to relaxation of the polariton density distribution in momentum space with the resultant formation of a non-equilibrium condensate manifested by a standing wave pattern across the whole sample.
\end{abstract}

\maketitle


\emph{Introduction.---} Self-organisation of nonlinear waves plays a fundamental role in a wide variety of phenomena, which in many cases have shaped the development of key areas of modern physics. These effects include Bose-Einstein condensation (BEC), spontaneous pattern formation, turbulence, solitons and topological defects. Solitons are self-sustained objects characterised by the energy localisation in space and time through a balance between nonlinearity and dispersion. They contain a broad spectrum of waves with different energies and momenta. By contrast BEC is characterised by a quasi-homogeneous density distribution in real space and by a narrow spectrum in momentum space. In optical systems condensates and solitons typically interact with background radiation. 

In nonlinear optics the interplay between nonlinearity, spatial, and temporal degrees of freedom is particularly interesting. It enables the study of ultra-broadband emission and multi-mode solitons~\cite{Wright2015} in multi-mode fibres and Bose-Einstein-like condensation of classical waves in nonlinear crystals~\cite{Sun2012}. Both effects arise from scattering between different transverse modes~\cite{Wright2015,Krupa2016}. Describing such complex systems analytically or numerically from a microscopic quantum point of view poses great challenges~\cite{Poletti2008}. One approach is based on kinetic wave theory and principles of thermodynamics~\cite{Connaughton2005}, which was also successfully used to explain supercontinuum generation in optical fibres~\cite{Barviau2009} and incoherent spectral solitons~\cite{Picozzi2008SpectralDomain,Gorbach2006Spectral-discreteSpace}. The second approach employs coupled nonlinear Schr\"{o}dinger equations, thus neglecting any incoherent wave population, which also has been used to describe multi-mode solitons~\cite{Sukhorukov2001Multi-solitonModes}.

Polaritons in optical microresonators, where strong exciton-photon hybridisation enables giant $\chi^{(3)}$ optical nonlinearity~\cite{Gippius2007, Walker2015}, form a unique laboratory for the study of nonlinear collective phenomena, including BEC and polariton lasing~\cite{Kasprzak2006,Galbiati2012, Bajoni2008, Sun2017}, self-organisation through multiple polariton-polariton scattering~\cite{Krizhanovskii2008}, quantised vortices~\cite{Lagoudakis2008QuatizedVortices,Tosi2012} and solitons ~\cite{Amo2011,Sich2011,Chana2015}. While in planar 2D microcavities polariton-polariton scattering usually occurs between the states residing in a single band formed by the lower polariton branch~\cite{Stevenson2000}, a range of scattering channels opens up in laterally confined systems, such as microcavity wires (MCWs) \cite{Ferrier2011,Wertz2010} where nonlinear interactions can mix between different transverse polariton modes~\cite{Dasbach2002}. Theoretically this mixing can lead to competition between modes of different parity and formation  of parity switching waves and parity solitons under static nonresonant excitation~\cite{Sigurdsson2015SwitchingCondensates,Sigurdsson2017ParityChannels}. At the same time, spatio-temporal kinetics of multi-mode nonlinear polariton pulses in this system so far remains unexplored. This article  fills this gap, addressing transitions between different polariton phases: from multi-mode wavepackets to solitons and, finally, to a nonequilibrium condensate. 

\begin{figure}
    \centering
    \includegraphics[width=\columnwidth]{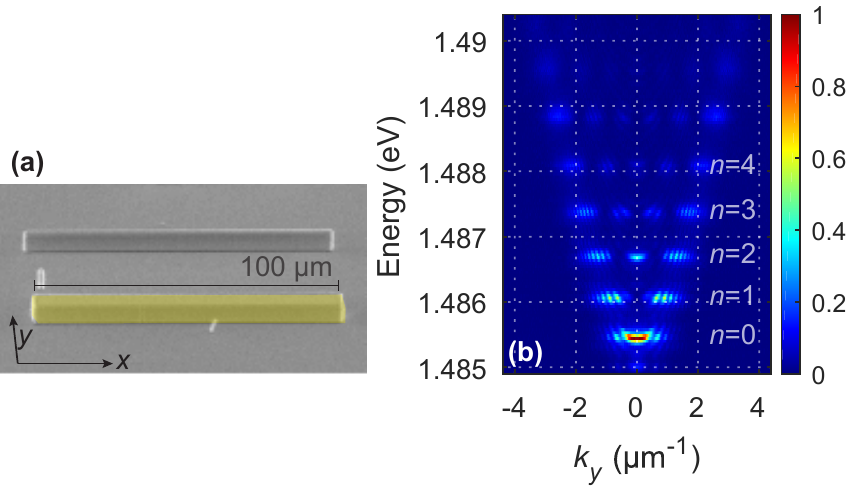}
    \caption{(a) SEM image of the sample with etched microwires. The \(8~\um\) by \(100~\um\) wire is shaded in yellow. (b) Energy-momentum dispersion of the lower polariton branch measured across the wire, along the \(y\)-axis, showing different energy modes arising from lateral photonic confinement. The fine modulation of the mode dispersions arises from interference due to reflection from the polished side of the substrate~\cite{Tartakovskii2000Polariton-polaritonCase}.}
    \label{fig:intro}
\end{figure}

We study nonlinear kinetic evolution of a multi-mode polariton system in a 100~$\mu$m-long MCW, which is excited with quasi-resonant pulses at a finite momentum. In the nonlinear regime polariton-polariton interactions redistribute the particles between several transverse lower polariton modes. Above a certain threshold, this process leads to a dominant occupation of the ground mode in a finite range of non-zero momenta close to the inflection point of the dispersion curve, where polariton effective mass becomes negative. This manifests itself as propagating bright single- and double-peak solitons. At even stronger excitation, cascading polariton-polariton and polariton-exciton scattering leads to relaxation of the polariton density to lower momenta, forming a non-equilibrium analogue of BEC, characterised by a standing wave pattern. It is possible to achieve this quasi-thermalised state because of the long polariton lifetime of $\simeq30$~ps, which allows significant redistribution of the polariton density since characteristic interaction times are much shorter then the lifetime. The data is in quantitative agreement with results of numerical modelling using the generalised Gross-Pitaevskii equation (see the Supplemental Materials~\cite{SM}). We note that previously formation of conservative bright polariton solitons have been reported in a narrow and long MCW~\cite{Skryabin2017BackwardWire} where, in contrast to the present work, predominantly only the ground polariton transverse mode was excited and no multi-mode evolution, mode competition, and standing wave condensation were observed.


\emph{Results.---} Our sample is a $\sfrac{3\uplambda}{2}$ microcavity with 3 InGaAs quantum wells  (QWs, 10~nm thick, 4\% Indium), and was previously described in the Ref.~\cite{Tinkler2015}. Distributed Bragg reflector (DBR) mirrors are GaAs/AlGaAs (85\% Al) with 26 repeats on the bottom mirror and 23 repeats on the top. The detuning of the ground, \(n=0\), photonic mode and the exciton is $\simeq-4.07$~meV with a Rabi splitting of $\simeq$~4.12~meV and exciton-polariton lifetime of \(\simeq 30\)~ps. The exciton emission was at $\simeq$~832~nm (\(\simeq1.49\)~eV). The top mirror was partially etched, defining \(100~\um\)-long mesas with a width of \(8~\um\) (Fig.~\ref{fig:intro}(a)). The lateral confinement of the photonic mode generates discrete energy levels labelled as \(n=0,1,2,...\) (where \(n\) is the number of nodes in the photon field distribution across the wire), which can be seen in the far-field polariton photo-luminescence (PL) under a low-power non-resonant excitation (Fig.~\ref{fig:intro}(b)).


\begin{figure}
    \centering
    \includegraphics[width=\columnwidth]{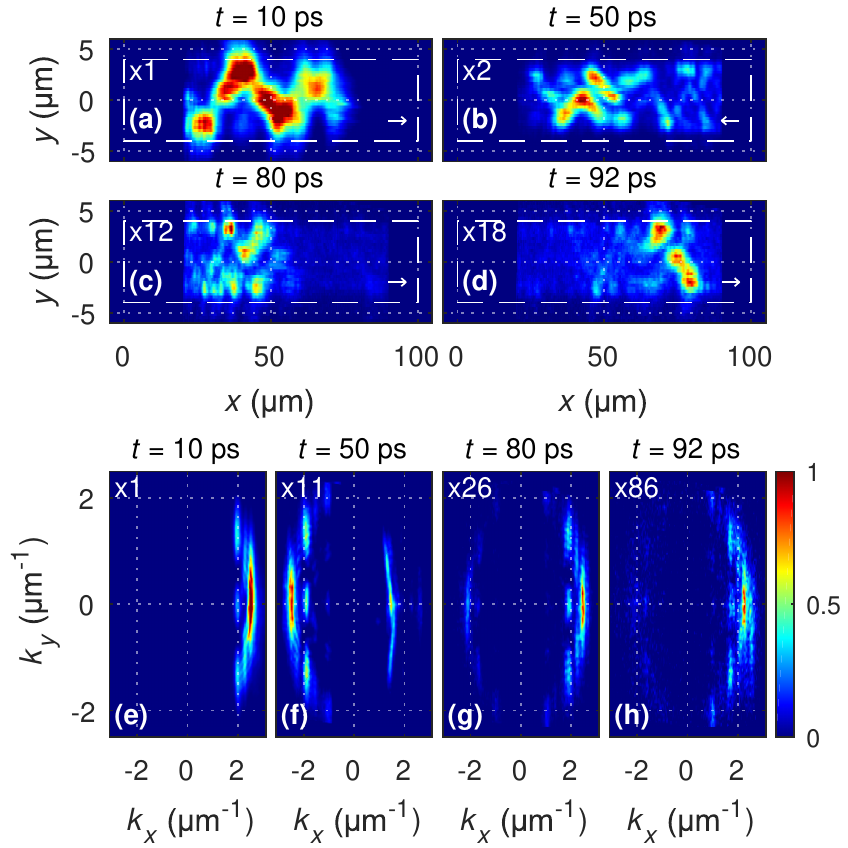}
    \caption{Low power, \(P_1=90~\mu\)W, emission characterisation. (a-d) Reconstructed real-space images of the polariton pulse propagating in the MCW at different times. White dashed rectangles show the outline of the MCW. Arrows in lower right corners indicate the direction of travel of the pulses. (e-h) are the corresponding snapshots of the momentum space at the same times as (a-d) respectively. All pseudo-colour scales are linear.}
    \label{fig:lowp}
\end{figure}

We applied a quasi-resonant pulsed excitation laser at an angle of incidence relative to the sample top surface corresponding to \(k_x\simeq 2.4~\ium\) and \(k_y\simeq0\).  The excitation beam was spectrally-filtered to approx. 5-7~ps duration FWHM (corresponding to \(\simeq0.3\)~meV energy width) and focused into a spot size of $\simeq 20~\um$ close to one end of the wire. The finite width of the pulse in momentum, \(\Delta k_x\simeq 0.4~\ium\), as well as Rayleigh scattering from the edges of the etched MCW enables efficient excitation of three (\(n=0, 1, 2\)) transverse lower polariton modes (Fig.~\ref{fig:lowp}(e)). We start with the lowest excitation power, \(P_1=90~\mu\)W (Fig.~\ref{fig:lowp}), when polariton-polariton interactions are negligible. The excited polariton modes have different group velocities in the range of \(\sim 1\) to \(\sim3~\um\)/ps, which, in addition to polariton group velocity dispersion of each transverse polariton mode, leads to spreading of the pulse in real-space. The interference between the transverse modes also results in a visible 'snaking' (as in the Ref.~\cite{Anton2013}) of the pulse in real space (Fig.~\ref{fig:lowp}(a)). 

The long polariton lifetime allows us to observe several cycles of the pulse moving back and forth along the wire. Fig.~\ref{fig:lowp} provides several snap-shots of this process showing the real-space images and the corresponding k-space distributions. Within $\simeq 30$~ps after the excitation, the front of the pulse quickly reaches the end of the wire, where it is elastically reflected backwards so that the momentum of polariton emission changes its sign (Fig.~\ref{fig:lowp}(f)). During reflections from the ends of the wire, polariton modes of higher orders, \emph{i.e.} \(n=3\) and \(4\), are also populated through the elastic scattering of the pulse from imperfections Figs.~\ref{fig:lowp}(f-h)). As a result, the interference between the transverse low- and high-order modes enhances the overall pulse spreading with time producing more complex real-space patterns (Figs.~\ref{fig:lowp}(b,c,d)). Overall, at the low pump power the momentum emission associated with different modes is almost the same at $\simeq 10$ and $\simeq 90$~ps, confirming low efficiency of polariton relaxation in momentum (and, hence, energy) space due to weak interactions with phonons, which is also reproduced in our modelling (see Fig.~S3 of the SM~\cite{SM}).


\begin{figure}
    \centering
    \includegraphics[width=\columnwidth]{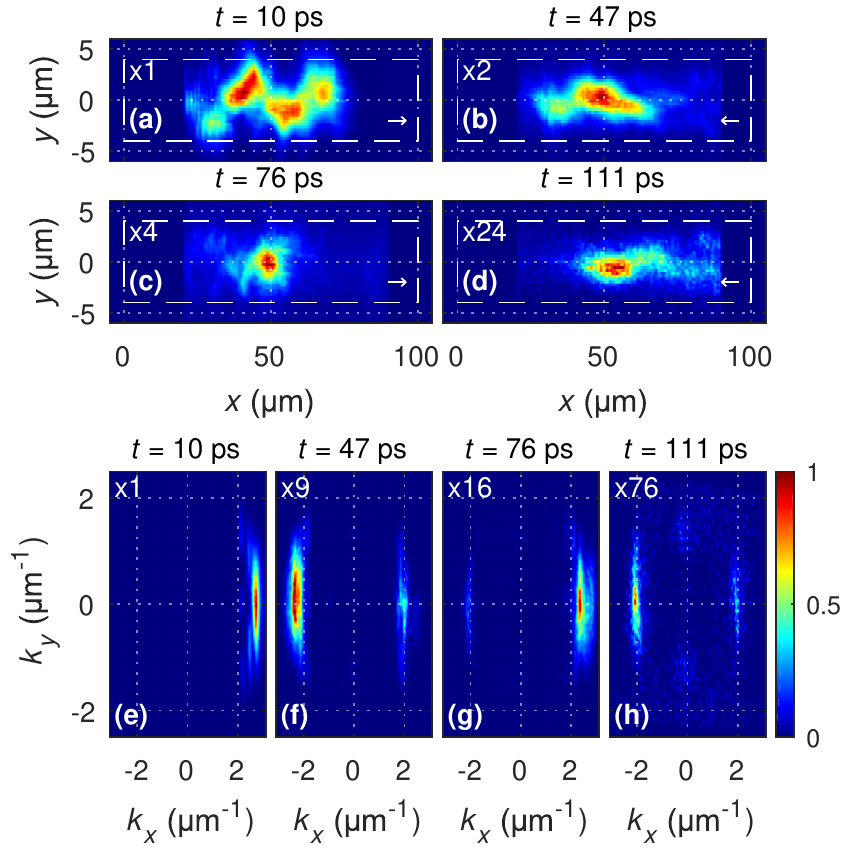}
    \caption{Medium power, \(P_2=540~\mu\)W, emission characterisation for different times. Labelling is the same as in Fig.~\ref{fig:lowp}}
    \label{fig:medp}
\end{figure}

At intermediate power, \(P_2=540~\mu\)W, the excitation k-vector plays a crucial role. Namely, since the point of inflection of the lower polariton mode (\(n=0\)) in this sample is at \(\simeq 2.1~\ium\), polaritons excited by the pump in mode $n=0$ have a negative effective mass. Hence, the interplay of the polariton group velocity dispersion with the repulsive polariton-polariton interactions and scattering can enable soliton formation~\cite{Sich2011}. Snap-shots of pulse evolution in real space and momentum space are shown in Fig~\ref{fig:medp}. The initial pulse propagation is very similar to the case of the low power, \(P_1\), as can be seen by comparing panels (a) and (b) in Figs.~\ref{fig:medp} and \ref{fig:lowp}. However, in contrast to the low power behaviour, here, at later times (50-80~ps), the polariton nonlinearity  results in the emergence of a single dominant mode, when individual energy levels can no longer be resolved in the momentum space (Figs.~\ref{fig:medp}(f,g), also Figs.~S4(c-g,j-k) of the SM), which coincides with a significant  narrowing of the pulse in real space (and hence in time) down to \(\simeq 10~\um\), as in Fig.~\ref{fig:medp}(c). The ratio of the peak polariton emission intensities at (50-80~ps) to that at 10~ps (Fig.~\ref{fig:medp}(f-h)) is $\simeq1.6$ times higher than the same ratio at the low excitation power (Figs.~\ref{fig:lowp}(f-h)), which is consistent with the concentration of pulse energy in the ground mode. Note, the observation of temporal nonlinear dynamics in our sample is enabled by the long polariton lifetime ($\simeq30$~ps), which is also significantly longer than the resolution of the streak camera (2~ps) used for the data acquisition.

\begin{figure}[t]
    \centering
    \includegraphics[width=\columnwidth]{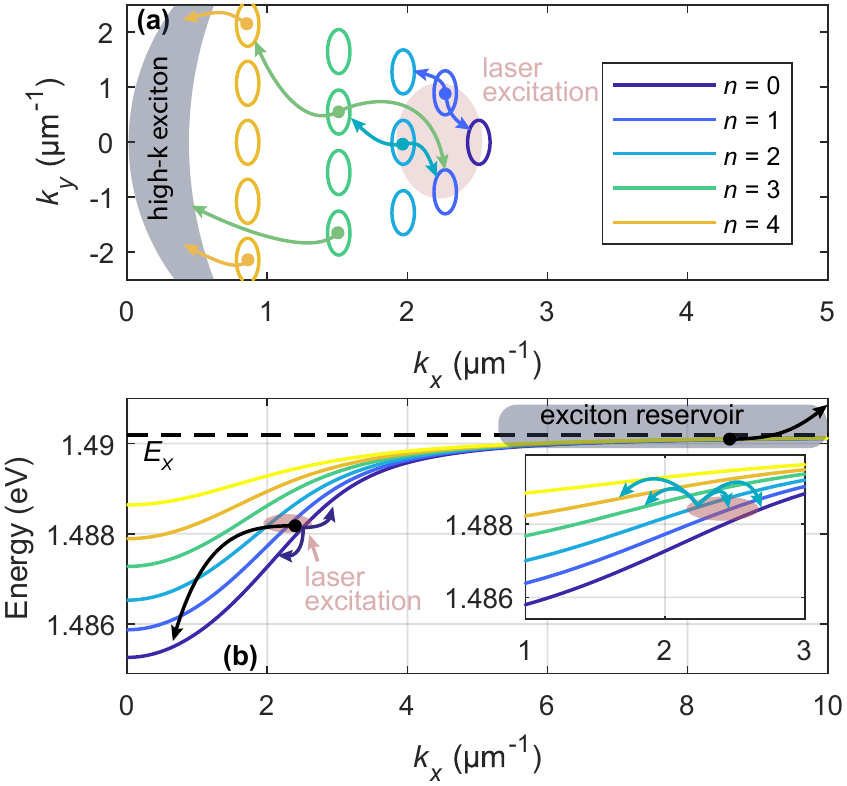}
    \caption{(a) Schematic of some of the possible combinations for \emph{inter-modal} polariton scattering in k-space at a fixed energy (\(\simeq1.486~\)eV); the coloured ovals approximate location and width of different lower polariton energy modes; (b) Schematic of \emph{intra-modal} polariton-polariton and polariton-exciton scattering leading to polariton relaxation. The inset corresponds to \emph{inter-modal} scattering shown in (a).}
    \label{fig:dsp}
\end{figure}

Kerr-like nonlinear interactions between transverse photonic modes in nonlinear crystals and optical fibres have been shown to lead to emergence of solitons and condensation of classical waves~\cite{Sun2012,Connaughton2005}. A similar process occurs in the polariton MCW where polaritons, excited within a certain momentum (energy) range, populate other initially empty polariton states through nonlinear polariton-polariton scattering. In turn, this maximises the population of the ground mode $n=0$ in the range of high momenta ($k\sim 2$-$2.5~\ium$). The interplay between negative polariton mass and nonlinear repulsive interactions between polaritons with different momenta in the ground mode leads to self-focusing and evolution of the system towards a temporal soliton at 50-75~ps. Some of the corresponding scattering channels are depicted in Fig.~\ref{fig:dsp}(a): interactions between polaritons residing initially in modes $n=1$ and $2$ result in a drastic increase of occupation in mode $n=0$ as well as the occupation of higher order modes ($n=3, 4$, and $5$). Furthermore, both inter-modal and intra-modal scattering spreads polariton population over a large range of k-vectors, thus minimising peak intensities in momentum space of the excited transverse ($n\geq1$) modes relative to the solitonic emission at $n=0$. Note that the polariton population (and hence nonlinearity) diminishes with time due to the finite lifetime, which together with the polariton dispersion leads to broadening of the polariton wavepacket at later times ($>75$~ps). The experimental results in Fig.~\ref{fig:medp}(f-h) are well reproduced by the numerical modelling taking into account coherent interaction between multiple transverse modes (see Fig.~S4 of the SM~\cite{SM})


\begin{figure}
    \centering
    \includegraphics[width=\columnwidth]{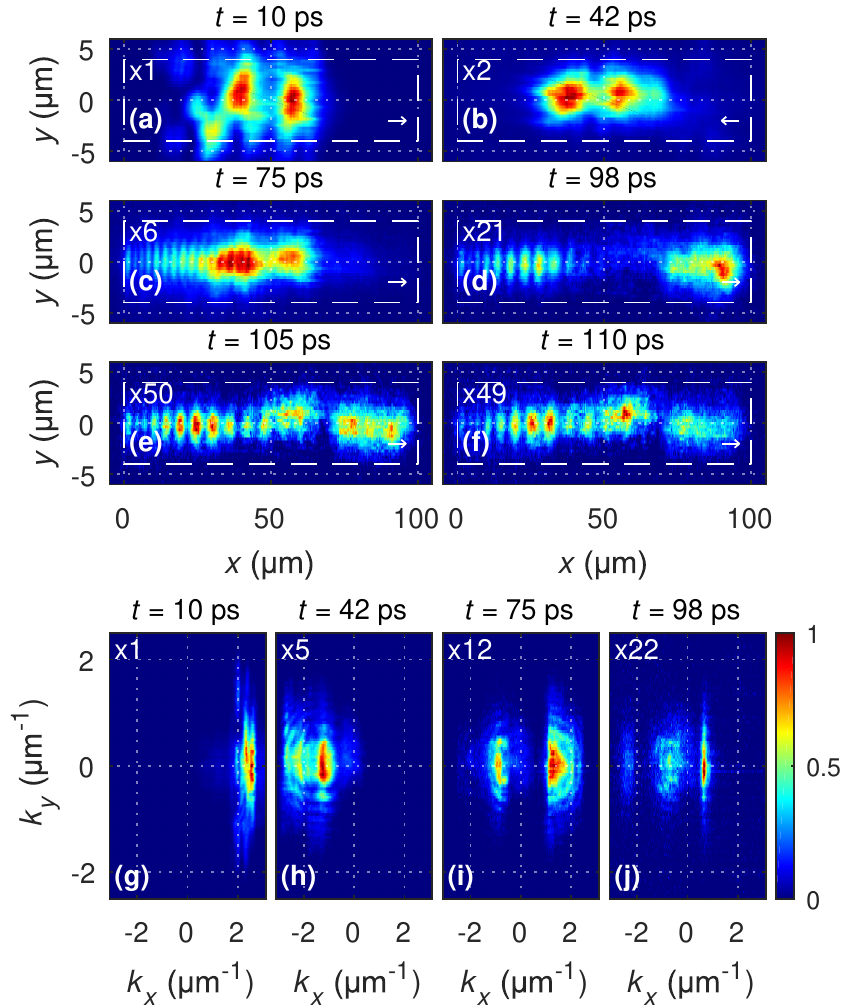}
    \caption{High power, \(P_3=800~\mu\)W, emission characterisation for different times. (a-f) Reconstructed real-space images of the polariton pulse propagating in the MCW at different times. White dashed rectangles show the outline of the MCW. Arrows in lower right corners indicate the direction of travel of the pulses. (g-j) are the corresponding snapshots of the momentum space at the same times as (a-d) respectively. All pseudo-colour scales are linear.}
    \label{fig:highp}
\end{figure}

The soliton regime described above does not correspond to a thermalised state, which is not possible to achieve at the intermediate excitation power due to the finite polariton lifetime. However, at a higher excitation power, thermalisation can speed up due to the increased rate of polariton-polariton scattering. At \(P_3=800~\mu\)W a soliton doublet \cite{Chana2015}, corresponding to the soliton fission regime, emerges already at 10-15~ps after the excitation and remains stable until $\simeq 75$~ps (Figs.~\ref{fig:highp}(b,c)). By 30-40~ps the emission acquires symmetry of the ground mode in momentum space along \(k_y\), and has the corresponding broad modulated spectrum arising from broadband inter-modal and intra-modal polariton-polariton scattering (modulation instability) as can be seen in Figs.~\ref{fig:highp}(h,i). A large part of the the soliton doublet spectrum now lies below the point of inflection (at $k_x\simeq1.8~\ium$), in the region where polariton effective mass is positive, and where wavepacket defocusing is expected. In this case, solitons can give up their energy to extended dispersive modes with lower k-vectors via Cherenkov radiation \cite{Hammani2010,Skryabin2003,Skryabin2017BackwardWire}. From a microscopical point of view, this process again can be understood as a result of multiple polariton-polariton scattering events. During each of these scatterings a pair of polaritons of the same energy scatters one to a lower and the second to a higher energy polariton states. This mechanism results in a gradual shift in momentum space of the maximum of polariton distribution with time to lower k-vectors since high-energy and high-k polaritons experience higher losses due to scattering to high-density high-momenta exciton-like states, the so-called exciton reservoir~\cite{Walker2017}. The losses may arise from interaction of polaritons with excitonic disorder~\cite{Krizhanovskii2001EnergyMicrocavities, Kulakovskii2002ImpactPolaritons} or polariton-phonon and polariton-electron scattering~\cite{Tartakovskii2003GiantTheory}. The mechanisms involving the reservoir are not directly taken into account in our numerical modelling (which reproduces experimental results well, see Figs.~S5 of the SM~\cite{SM}), but are accounted for phenomenologically by introducing excitonic decay rates higher than photonic. Furthermore, even though the energy of the lower polariton states is below that of bare uncoupled excitons, coherent pair polariton-polariton scattering may also effectively populate the latter mainly due to a very high density of exciton states (Fig.~\ref{fig:dsp}(a)). This is confirmed by our simulations (see Figs.~S5(j-l) of the SM~\cite{SM}). Finally, note, that polariton scattering with high-momenta excitons shown in Fig.~\ref{fig:dsp}(b) probably also plays an important role in the polariton relaxation~\cite{Savvidis2002RingMicrocavities}. Therefore, a number of mechanisms are potentially involved in spectral redistribution of polaritons in the wavepacket.

At $\simeq 75$ ps slow counter propagating waves emerge in the wire at \(k_x\simeq\pm0.8\)-\(1~\ium\) leading to formation of a modulated tail behind the doublet. At $\simeq100$~ps the polariton emission mostly peaks at $k_x\simeq 0.5$-$0.7~\ium$, lower than the momentum of the excitation pulse. This corresponds to onset of a standing wave with 17 maxima seen in Figs.~\ref{fig:highp}(d-f). The same effect is also observed in our modelling in Figs.~S5(d,h,j) of the SM~\cite{SM}. This standing wave arises from the interference across the whole wire between two waves at \(k_x\simeq\pm0.5\)-\(0.6~\ium\) which are long-range and coherent and hence form a macroscopically occupied state (a dynamic analogue of a nonequilibrium BEC). 


\emph{Discussion.---} 
We investigated temporal evolution of multi-mode polariton wavepackets excited with a finite momentum in a long microcavity wire resonator. Polariton nonlinearity gives rise to inter-modal and intra-modal polariton-polariton scattering, and results in evolution towards a quasi-thermalised polariton gas in a succession of different phases: multi-mode wavepacket \textrightarrow~soliton \textrightarrow~soliton doublet \textrightarrow~dynamic condensate. 

Our findings show that in a sample with a long polariton lifetime, polariton condensates can emerge out of a resonantly excited polariton cloud. By varying the energy, the bandwidth, and the power of the excitation pulse it is possible to control the excited polariton modes and their energy-momentum distributions. Resonant excitation can also allow the control of the spin degree of freedom, which can be useful for investigation of BKT phases~\cite{Caputo2016} associated with the emergence of half- or full-spin vortex excitations in polariton system with spin-dependent anisotropy of polariton-polariton interactions~\cite{Vladimirova2010}, so far a completely unexplored field. 

\begin{acknowledgments}
\emph{Acknowledgements.---} 
MS and DNK acknowledge the support from the Leverhulme Trust grant No. RPG-2013-339. MS, JKC, PMW, BR, MSS, and DNK acknowledge the support from the EPSRC grants EP/J007544/1, EP/N031776/1, and the ERC Advanced Grant EXCIPOL 320570. DVS acknowledges Russian Foundation for Basic Research (16-52-150006);  ITMO University Fellowship through the Government of Russia grant 074-U01. HS and IAS acknowledge the support by the Research Fund of the University of Iceland, The Icelandic Research Fund, Grant No. 163082-051 and the Project 3.2614.2017/4.6 of the Ministry of Education and Science of Russian Federation. IAS, MSS and DNK from Megagrant No. 14.Y26.31.0015 of the Ministry of Education and Science of Russian Federation.

We thank Marzena Szyma\'{n}ska for helpful discussions.
\end{acknowledgments}


\bibliography{Mendeley_Refs_for_Kinetic_Condensation_Paper.bib}

\begin{thebibliography}{41}%
\makeatletter
\providecommand \@ifxundefined [1]{%
 \@ifx{#1\undefined}
}%
\providecommand \@ifnum [1]{%
 \ifnum #1\expandafter \@firstoftwo
 \else \expandafter \@secondoftwo
 \fi
}%
\providecommand \@ifx [1]{%
 \ifx #1\expandafter \@firstoftwo
 \else \expandafter \@secondoftwo
 \fi
}%
\providecommand \natexlab [1]{#1}%
\providecommand \enquote  [1]{``#1''}%
\providecommand \bibnamefont  [1]{#1}%
\providecommand \bibfnamefont [1]{#1}%
\providecommand \citenamefont [1]{#1}%
\providecommand \href@noop [0]{\@secondoftwo}%
\providecommand \href [0]{\begingroup \@sanitize@url \@href}%
\providecommand \@href[1]{\@@startlink{#1}\@@href}%
\providecommand \@@href[1]{\endgroup#1\@@endlink}%
\providecommand \@sanitize@url [0]{\catcode `\\12\catcode `\$12\catcode
  `\&12\catcode `\#12\catcode `\^12\catcode `\_12\catcode `\%12\relax}%
\providecommand \@@startlink[1]{}%
\providecommand \@@endlink[0]{}%
\providecommand \url  [0]{\begingroup\@sanitize@url \@url }%
\providecommand \@url [1]{\endgroup\@href {#1}{\urlprefix }}%
\providecommand \urlprefix  [0]{URL }%
\providecommand \Eprint [0]{\href }%
\providecommand \doibase [0]{http://dx.doi.org/}%
\providecommand \selectlanguage [0]{\@gobble}%
\providecommand \bibinfo  [0]{\@secondoftwo}%
\providecommand \bibfield  [0]{\@secondoftwo}%
\providecommand \translation [1]{[#1]}%
\providecommand \BibitemOpen [0]{}%
\providecommand \bibitemStop [0]{}%
\providecommand \bibitemNoStop [0]{.\EOS\space}%
\providecommand \EOS [0]{\spacefactor3000\relax}%
\providecommand \BibitemShut  [1]{\csname bibitem#1\endcsname}%
\let\auto@bib@innerbib\@empty
\bibitem [{\citenamefont {Wright}\ \emph {et~al.}(2015)\citenamefont {Wright},
  \citenamefont {Wabnitz}, \citenamefont {Christodoulides},\ and\ \citenamefont
  {Wise}}]{Wright2015}%
  \BibitemOpen
  \bibfield  {author} {\bibinfo {author} {\bibfnamefont {L.~G.}\ \bibnamefont
  {Wright}}, \bibinfo {author} {\bibfnamefont {S.}~\bibnamefont {Wabnitz}},
  \bibinfo {author} {\bibfnamefont {D.N.}\ \bibnamefont {Christodoulides}}, \
  and\ \bibinfo {author} {\bibfnamefont {F.~W.}\ \bibnamefont {Wise}},\
  }\bibfield  {title} {\enquote {\bibinfo {title} {{Ultrabroadband dispersive
  radiation by spatiotemporal oscillation of multimode waves}},}\ }\href
  {\doibase 10.1103/PhysRevLett.115.223902} {\bibfield  {journal} {\bibinfo
  {journal} {Physical Review Letters}\ }\textbf {\bibinfo {volume} {115}},\
  \bibinfo {pages} {223902} (\bibinfo {year} {2015})}\BibitemShut {NoStop}%
\bibitem [{\citenamefont {Sun}\ \emph {et~al.}(2012)\citenamefont {Sun},
  \citenamefont {Jia}, \citenamefont {Barsi}, \citenamefont {Rica},
  \citenamefont {Picozzi},\ and\ \citenamefont {Fleischer}}]{Sun2012}%
  \BibitemOpen
  \bibfield  {author} {\bibinfo {author} {\bibfnamefont {C.}~\bibnamefont
  {Sun}}, \bibinfo {author} {\bibfnamefont {S.}~\bibnamefont {Jia}}, \bibinfo
  {author} {\bibfnamefont {C.}~\bibnamefont {Barsi}}, \bibinfo {author}
  {\bibfnamefont {S.}~\bibnamefont {Rica}}, \bibinfo {author} {\bibfnamefont
  {A.}~\bibnamefont {Picozzi}}, \ and\ \bibinfo {author} {\bibfnamefont
  {J.~W.}\ \bibnamefont {Fleischer}},\ }\bibfield  {title} {\enquote {\bibinfo
  {title} {{Observation of the kinetic condensation of classical waves}},}\
  }\href {\doibase 10.1038/nphys2278} {\bibfield  {journal} {\bibinfo
  {journal} {Nature Physics}\ }\textbf {\bibinfo {volume} {8}},\ \bibinfo
  {pages} {470--474} (\bibinfo {year} {2012})}\BibitemShut {NoStop}%
\bibitem [{\citenamefont {Krupa}\ \emph {et~al.}(2016)\citenamefont {Krupa},
  \citenamefont {Tonello}, \citenamefont {Barth{\'{e}}l{\'{e}}my},
  \citenamefont {Couderc}, \citenamefont {Shalaby}, \citenamefont {Bendahmane},
  \citenamefont {Millot},\ and\ \citenamefont {Wabnitz}}]{Krupa2016}%
  \BibitemOpen
  \bibfield  {author} {\bibinfo {author} {\bibfnamefont {K.}~\bibnamefont
  {Krupa}}, \bibinfo {author} {\bibfnamefont {A.}~\bibnamefont {Tonello}},
  \bibinfo {author} {\bibfnamefont {A.}~\bibnamefont {Barth{\'{e}}l{\'{e}}my}},
  \bibinfo {author} {\bibfnamefont {V.}~\bibnamefont {Couderc}}, \bibinfo
  {author} {\bibfnamefont {B.~M.}\ \bibnamefont {Shalaby}}, \bibinfo {author}
  {\bibfnamefont {A.}~\bibnamefont {Bendahmane}}, \bibinfo {author}
  {\bibfnamefont {G.}~\bibnamefont {Millot}}, \ and\ \bibinfo {author}
  {\bibfnamefont {S.}~\bibnamefont {Wabnitz}},\ }\bibfield  {title} {\enquote
  {\bibinfo {title} {{Observation of geometric parametric instability induced
  by the periodic spatial self-imaging of multimode waves}},}\ }\href {\doibase
  10.1103/PhysRevLett.116.183901} {\bibfield  {journal} {\bibinfo  {journal}
  {Physical Review Letters}\ }\textbf {\bibinfo {volume} {116}},\ \bibinfo
  {pages} {183901} (\bibinfo {year} {2016})}\BibitemShut {NoStop}%
\bibitem [{\citenamefont {Poletti}\ and\ \citenamefont
  {Horak}(2008)}]{Poletti2008}%
  \BibitemOpen
  \bibfield  {author} {\bibinfo {author} {\bibfnamefont {F.}~\bibnamefont
  {Poletti}}\ and\ \bibinfo {author} {\bibfnamefont {P.}~\bibnamefont
  {Horak}},\ }\bibfield  {title} {\enquote {\bibinfo {title} {{Description of
  ultrashort pulse propagation in multimode optical fibers}},}\ }\href
  {\doibase 10.1364/JOSAB.25.001645} {\bibfield  {journal} {\bibinfo  {journal}
  {Journal of the Optical Society of America B}\ }\textbf {\bibinfo {volume}
  {25}},\ \bibinfo {pages} {1645} (\bibinfo {year} {2008})}\BibitemShut
  {NoStop}%
\bibitem [{\citenamefont {Connaughton}\ \emph {et~al.}(2005)\citenamefont
  {Connaughton}, \citenamefont {Josserand}, \citenamefont {Picozzi},
  \citenamefont {Pomeau},\ and\ \citenamefont {Rica}}]{Connaughton2005}%
  \BibitemOpen
  \bibfield  {author} {\bibinfo {author} {\bibfnamefont {C.}~\bibnamefont
  {Connaughton}}, \bibinfo {author} {\bibfnamefont {C.}~\bibnamefont
  {Josserand}}, \bibinfo {author} {\bibfnamefont {A.}~\bibnamefont {Picozzi}},
  \bibinfo {author} {\bibfnamefont {Y.}~\bibnamefont {Pomeau}}, \ and\ \bibinfo
  {author} {\bibfnamefont {S.}~\bibnamefont {Rica}},\ }\bibfield  {title}
  {\enquote {\bibinfo {title} {{Condensation of classical nonlinear waves}},}\
  }\href {\doibase 10.1103/PhysRevLett.95.263901} {\bibfield  {journal}
  {\bibinfo  {journal} {Physical Review Letters}\ }\textbf {\bibinfo {volume}
  {95}},\ \bibinfo {pages} {263901} (\bibinfo {year} {2005})}\BibitemShut
  {NoStop}%
\bibitem [{\citenamefont {Barviau}\ \emph {et~al.}(2009)\citenamefont
  {Barviau}, \citenamefont {Kibler}, \citenamefont {Kudlinski}, \citenamefont
  {Mussot}, \citenamefont {Millot},\ and\ \citenamefont
  {Picozzi}}]{Barviau2009}%
  \BibitemOpen
  \bibfield  {author} {\bibinfo {author} {\bibfnamefont {B.}~\bibnamefont
  {Barviau}}, \bibinfo {author} {\bibfnamefont {B.}~\bibnamefont {Kibler}},
  \bibinfo {author} {\bibfnamefont {A.}~\bibnamefont {Kudlinski}}, \bibinfo
  {author} {\bibfnamefont {A.}~\bibnamefont {Mussot}}, \bibinfo {author}
  {\bibfnamefont {G.}~\bibnamefont {Millot}}, \ and\ \bibinfo {author}
  {\bibfnamefont {A.}~\bibnamefont {Picozzi}},\ }\bibfield  {title} {\enquote
  {\bibinfo {title} {{Experimental signature of optical wave thermalization
  through supercontinuum generation in photonic crystal fiber}},}\ }\href
  {\doibase 10.1364/OE.17.007392} {\bibfield  {journal} {\bibinfo  {journal}
  {Optics Express}\ }\textbf {\bibinfo {volume} {17}},\ \bibinfo {pages} {7392}
  (\bibinfo {year} {2009})}\BibitemShut {NoStop}%
\bibitem [{\citenamefont {Picozzi}\ \emph {et~al.}(2008)\citenamefont
  {Picozzi}, \citenamefont {Pitois},\ and\ \citenamefont
  {Millot}}]{Picozzi2008SpectralDomain}%
  \BibitemOpen
  \bibfield  {author} {\bibinfo {author} {\bibfnamefont {A.}~\bibnamefont
  {Picozzi}}, \bibinfo {author} {\bibfnamefont {S.}~\bibnamefont {Pitois}}, \
  and\ \bibinfo {author} {\bibfnamefont {G.}~\bibnamefont {Millot}},\
  }\bibfield  {title} {\enquote {\bibinfo {title} {{Spectral incoherent
  solitons: a localized soliton behavior in the frequency domain.}}}\ }\href
  {\doibase 10.1103/PhysRevLett.101.093901} {\bibfield  {journal} {\bibinfo
  {journal} {Physical Review Letters}\ }\textbf {\bibinfo {volume} {101}},\
  \bibinfo {pages} {093901} (\bibinfo {year} {2008})}\BibitemShut {NoStop}%
\bibitem [{\citenamefont {Gorbach}\ and\ \citenamefont
  {Skryabin}(2006)}]{Gorbach2006Spectral-discreteSpace}%
  \BibitemOpen
  \bibfield  {author} {\bibinfo {author} {\bibfnamefont {A.~V.}\ \bibnamefont
  {Gorbach}}\ and\ \bibinfo {author} {\bibfnamefont {D.~V.}\ \bibnamefont
  {Skryabin}},\ }\bibfield  {title} {\enquote {\bibinfo {title}
  {{Spectral-discrete solitons and localization in frequency space}},}\ }\href
  {\doibase 10.1364/OL.31.003309} {\bibfield  {journal} {\bibinfo  {journal}
  {Optics Letters}\ }\textbf {\bibinfo {volume} {31}},\ \bibinfo {pages} {3309}
  (\bibinfo {year} {2006})}\BibitemShut {NoStop}%
\bibitem [{\citenamefont {Sukhorukov}\ \emph {et~al.}(2001)\citenamefont
  {Sukhorukov}, \citenamefont {Ankiewicz},\ and\ \citenamefont
  {Akhmediev}}]{Sukhorukov2001Multi-solitonModes}%
  \BibitemOpen
  \bibfield  {author} {\bibinfo {author} {\bibfnamefont {A.~A.}\ \bibnamefont
  {Sukhorukov}}, \bibinfo {author} {\bibfnamefont {A.}~\bibnamefont
  {Ankiewicz}}, \ and\ \bibinfo {author} {\bibfnamefont {N.~N.}\ \bibnamefont
  {Akhmediev}},\ }\bibfield  {title} {\enquote {\bibinfo {title}
  {{Multi-soliton complexes in a sea of radiation modes}},}\ }\href {\doibase
  10.1016/S0030-4018(01)01336-0} {\bibfield  {journal} {\bibinfo  {journal}
  {Optics Communications}\ }\textbf {\bibinfo {volume} {195}},\ \bibinfo
  {pages} {293--302} (\bibinfo {year} {2001})}\BibitemShut {NoStop}%
\bibitem [{\citenamefont {Gippius}\ \emph {et~al.}(2007)\citenamefont
  {Gippius}, \citenamefont {Shelykh}, \citenamefont {Solnyshkov}, \citenamefont
  {Gavrilov}, \citenamefont {Rubo}, \citenamefont {Kavokin}, \citenamefont
  {Tikhodeev},\ and\ \citenamefont {Malpuech}}]{Gippius2007}%
  \BibitemOpen
  \bibfield  {author} {\bibinfo {author} {\bibfnamefont {N.~A.}\ \bibnamefont
  {Gippius}}, \bibinfo {author} {\bibfnamefont {I.~A.}\ \bibnamefont
  {Shelykh}}, \bibinfo {author} {\bibfnamefont {D.~D.}\ \bibnamefont
  {Solnyshkov}}, \bibinfo {author} {\bibfnamefont {S.~S.}\ \bibnamefont
  {Gavrilov}}, \bibinfo {author} {\bibfnamefont {Y.~G.}\ \bibnamefont {Rubo}},
  \bibinfo {author} {\bibfnamefont {A.~V.}\ \bibnamefont {Kavokin}}, \bibinfo
  {author} {\bibfnamefont {S.~G.}\ \bibnamefont {Tikhodeev}}, \ and\ \bibinfo
  {author} {\bibfnamefont {G.}~\bibnamefont {Malpuech}},\ }\bibfield  {title}
  {\enquote {\bibinfo {title} {{Polarization multistability of cavity
  polaritons}},}\ }\href {\doibase 10.1103/PhysRevLett.98.236401} {\bibfield
  {journal} {\bibinfo  {journal} {Physical Review Letters}\ }\textbf {\bibinfo
  {volume} {98}},\ \bibinfo {pages} {236401} (\bibinfo {year}
  {2007})}\BibitemShut {NoStop}%
\bibitem [{\citenamefont {Walker}\ \emph {et~al.}(2015)\citenamefont {Walker},
  \citenamefont {Tinkler}, \citenamefont {Skryabin}, \citenamefont {Yulin},
  \citenamefont {Royall}, \citenamefont {Farrer}, \citenamefont {Ritchie},
  \citenamefont {Skolnick},\ and\ \citenamefont {Krizhanovskii}}]{Walker2015}%
  \BibitemOpen
  \bibfield  {author} {\bibinfo {author} {\bibfnamefont {P.~M.}\ \bibnamefont
  {Walker}}, \bibinfo {author} {\bibfnamefont {L.}~\bibnamefont {Tinkler}},
  \bibinfo {author} {\bibfnamefont {D.~V.}\ \bibnamefont {Skryabin}}, \bibinfo
  {author} {\bibfnamefont {A.}~\bibnamefont {Yulin}}, \bibinfo {author}
  {\bibfnamefont {B.}~\bibnamefont {Royall}}, \bibinfo {author} {\bibfnamefont
  {I.}~\bibnamefont {Farrer}}, \bibinfo {author} {\bibfnamefont {D.~A.}\
  \bibnamefont {Ritchie}}, \bibinfo {author} {\bibfnamefont {M.~S.}\
  \bibnamefont {Skolnick}}, \ and\ \bibinfo {author} {\bibfnamefont {D.~N.}\
  \bibnamefont {Krizhanovskii}},\ }\bibfield  {title} {\enquote {\bibinfo
  {title} {{Ultra-low-power hybrid light–matter solitons}},}\ }\href
  {\doibase 10.1038/ncomms9317} {\bibfield  {journal} {\bibinfo  {journal}
  {Nature Communications}\ }\textbf {\bibinfo {volume} {6}},\ \bibinfo {pages}
  {8317} (\bibinfo {year} {2015})}\BibitemShut {NoStop}%
\bibitem [{\citenamefont {Kasprzak}\ \emph {et~al.}(2006)\citenamefont
  {Kasprzak}, \citenamefont {Richard}, \citenamefont {Kundermann},
  \citenamefont {Baas}, \citenamefont {Jeambrun}, \citenamefont {Keeling},
  \citenamefont {Marchetti}, \citenamefont {Szyma{\'{n}}ska}, \citenamefont
  {Andr{\'{e}}}, \citenamefont {Staehli}, \citenamefont {Savona}, \citenamefont
  {Littlewood}, \citenamefont {Deveaud},\ and\ \citenamefont
  {Dang}}]{Kasprzak2006}%
  \BibitemOpen
  \bibfield  {author} {\bibinfo {author} {\bibfnamefont {J.}~\bibnamefont
  {Kasprzak}}, \bibinfo {author} {\bibfnamefont {M.}~\bibnamefont {Richard}},
  \bibinfo {author} {\bibfnamefont {S.}~\bibnamefont {Kundermann}}, \bibinfo
  {author} {\bibfnamefont {A.}~\bibnamefont {Baas}}, \bibinfo {author}
  {\bibfnamefont {P.}~\bibnamefont {Jeambrun}}, \bibinfo {author}
  {\bibfnamefont {J.~M.~J.}\ \bibnamefont {Keeling}}, \bibinfo {author}
  {\bibfnamefont {F.~M.}\ \bibnamefont {Marchetti}}, \bibinfo {author}
  {\bibfnamefont {M.~H.}\ \bibnamefont {Szyma{\'{n}}ska}}, \bibinfo {author}
  {\bibfnamefont {R.}~\bibnamefont {Andr{\'{e}}}}, \bibinfo {author}
  {\bibfnamefont {J.~L.}\ \bibnamefont {Staehli}}, \bibinfo {author}
  {\bibfnamefont {V.}~\bibnamefont {Savona}}, \bibinfo {author} {\bibfnamefont
  {P.~B.}\ \bibnamefont {Littlewood}}, \bibinfo {author} {\bibfnamefont
  {B.}~\bibnamefont {Deveaud}}, \ and\ \bibinfo {author} {\bibfnamefont
  {Le~Si}\ \bibnamefont {Dang}},\ }\bibfield  {title} {\enquote {\bibinfo
  {title} {{Bose-Einstein condensation of exciton polaritons}},}\ }\href
  {\doibase 10.1038/nature05131} {\bibfield  {journal} {\bibinfo  {journal}
  {Nature}\ }\textbf {\bibinfo {volume} {443}},\ \bibinfo {pages} {409--414}
  (\bibinfo {year} {2006})}\BibitemShut {NoStop}%
\bibitem [{\citenamefont {Galbiati}\ \emph {et~al.}(2012)\citenamefont
  {Galbiati}, \citenamefont {Ferrier}, \citenamefont {Solnyshkov},
  \citenamefont {Tanese}, \citenamefont {Wertz}, \citenamefont {Amo},
  \citenamefont {Abbarchi}, \citenamefont {Senellart}, \citenamefont {Sagnes},
  \citenamefont {Lema{\^{i}}tre}, \citenamefont {Galopin}, \citenamefont
  {Malpuech},\ and\ \citenamefont {Bloch}}]{Galbiati2012}%
  \BibitemOpen
  \bibfield  {author} {\bibinfo {author} {\bibfnamefont {M.}~\bibnamefont
  {Galbiati}}, \bibinfo {author} {\bibfnamefont {L.}~\bibnamefont {Ferrier}},
  \bibinfo {author} {\bibfnamefont {D.~D.}\ \bibnamefont {Solnyshkov}},
  \bibinfo {author} {\bibfnamefont {D.}~\bibnamefont {Tanese}}, \bibinfo
  {author} {\bibfnamefont {E.}~\bibnamefont {Wertz}}, \bibinfo {author}
  {\bibfnamefont {A.}~\bibnamefont {Amo}}, \bibinfo {author} {\bibfnamefont
  {M.}~\bibnamefont {Abbarchi}}, \bibinfo {author} {\bibfnamefont
  {P.}~\bibnamefont {Senellart}}, \bibinfo {author} {\bibfnamefont
  {I.}~\bibnamefont {Sagnes}}, \bibinfo {author} {\bibfnamefont
  {A.}~\bibnamefont {Lema{\^{i}}tre}}, \bibinfo {author} {\bibfnamefont
  {E.}~\bibnamefont {Galopin}}, \bibinfo {author} {\bibfnamefont
  {G.}~\bibnamefont {Malpuech}}, \ and\ \bibinfo {author} {\bibfnamefont
  {J.}~\bibnamefont {Bloch}},\ }\bibfield  {title} {\enquote {\bibinfo {title}
  {{Polariton condensation in photonic molecules}},}\ }\href {\doibase
  10.1103/PhysRevLett.108.126403} {\bibfield  {journal} {\bibinfo  {journal}
  {Physical Review Letters}\ }\textbf {\bibinfo {volume} {108}},\ \bibinfo
  {pages} {126403} (\bibinfo {year} {2012})}\BibitemShut {NoStop}%
\bibitem [{\citenamefont {Bajoni}\ \emph {et~al.}(2008)\citenamefont {Bajoni},
  \citenamefont {Senellart}, \citenamefont {Wertz}, \citenamefont {Sagnes},
  \citenamefont {Miard}, \citenamefont {Lema{\^{i}}tre},\ and\ \citenamefont
  {Bloch}}]{Bajoni2008}%
  \BibitemOpen
  \bibfield  {author} {\bibinfo {author} {\bibfnamefont {D.}~\bibnamefont
  {Bajoni}}, \bibinfo {author} {\bibfnamefont {P.}~\bibnamefont {Senellart}},
  \bibinfo {author} {\bibfnamefont {E.}~\bibnamefont {Wertz}}, \bibinfo
  {author} {\bibfnamefont {I.}~\bibnamefont {Sagnes}}, \bibinfo {author}
  {\bibfnamefont {A.}~\bibnamefont {Miard}}, \bibinfo {author} {\bibfnamefont
  {A.}~\bibnamefont {Lema{\^{i}}tre}}, \ and\ \bibinfo {author} {\bibfnamefont
  {J.}~\bibnamefont {Bloch}},\ }\bibfield  {title} {\enquote {\bibinfo {title}
  {{Polariton laser using single micropillar GaAs - GaAlAs semiconductor
  cavities}},}\ }\href {\doibase 10.1103/PhysRevLett.100.047401} {\bibfield
  {journal} {\bibinfo  {journal} {Physical Review Letters}\ }\textbf {\bibinfo
  {volume} {100}},\ \bibinfo {pages} {047401} (\bibinfo {year}
  {2008})}\BibitemShut {NoStop}%
\bibitem [{\citenamefont {Sun}\ \emph {et~al.}(2017)\citenamefont {Sun},
  \citenamefont {Wen}, \citenamefont {Yoon}, \citenamefont {Liu}, \citenamefont
  {Steger}, \citenamefont {Pfeiffer}, \citenamefont {West}, \citenamefont
  {Snoke},\ and\ \citenamefont {Nelson}}]{Sun2017}%
  \BibitemOpen
  \bibfield  {author} {\bibinfo {author} {\bibfnamefont {Y.}~\bibnamefont
  {Sun}}, \bibinfo {author} {\bibfnamefont {P.}~\bibnamefont {Wen}}, \bibinfo
  {author} {\bibfnamefont {Y.}~\bibnamefont {Yoon}}, \bibinfo {author}
  {\bibfnamefont {G.}~\bibnamefont {Liu}}, \bibinfo {author} {\bibfnamefont
  {M.}~\bibnamefont {Steger}}, \bibinfo {author} {\bibfnamefont {L.~N.}\
  \bibnamefont {Pfeiffer}}, \bibinfo {author} {\bibfnamefont {K.}~\bibnamefont
  {West}}, \bibinfo {author} {\bibfnamefont {D.~W.}\ \bibnamefont {Snoke}}, \
  and\ \bibinfo {author} {\bibfnamefont {K.~A.}\ \bibnamefont {Nelson}},\
  }\bibfield  {title} {\enquote {\bibinfo {title} {{Bose-Einstein condensation
  of long-lifetime polaritons in thermal equilibrium}},}\ }\href {\doibase
  10.1103/PhysRevLett.118.016602} {\bibfield  {journal} {\bibinfo  {journal}
  {Physical Review Letters}\ }\textbf {\bibinfo {volume} {118}},\ \bibinfo
  {pages} {016602} (\bibinfo {year} {2017})}\BibitemShut {NoStop}%
\bibitem [{\citenamefont {Krizhanovskii}\ \emph {et~al.}(2008)\citenamefont
  {Krizhanovskii}, \citenamefont {Gavrilov}, \citenamefont {Love},
  \citenamefont {Sanvitto}, \citenamefont {Gippius}, \citenamefont {Tikhodeev},
  \citenamefont {Kulakovskii}, \citenamefont {Whittaker}, \citenamefont
  {Skolnick},\ and\ \citenamefont {Roberts}}]{Krizhanovskii2008}%
  \BibitemOpen
  \bibfield  {author} {\bibinfo {author} {\bibfnamefont {D.~N.}\ \bibnamefont
  {Krizhanovskii}}, \bibinfo {author} {\bibfnamefont {S.~S.}\ \bibnamefont
  {Gavrilov}}, \bibinfo {author} {\bibfnamefont {A.~P~.D.}\ \bibnamefont
  {Love}}, \bibinfo {author} {\bibfnamefont {D.}~\bibnamefont {Sanvitto}},
  \bibinfo {author} {\bibfnamefont {N.~A.}\ \bibnamefont {Gippius}}, \bibinfo
  {author} {\bibfnamefont {S.~G.}\ \bibnamefont {Tikhodeev}}, \bibinfo {author}
  {\bibfnamefont {V.~D.}\ \bibnamefont {Kulakovskii}}, \bibinfo {author}
  {\bibfnamefont {D.~M.}\ \bibnamefont {Whittaker}}, \bibinfo {author}
  {\bibfnamefont {M.~S.}\ \bibnamefont {Skolnick}}, \ and\ \bibinfo {author}
  {\bibfnamefont {J.~S.}\ \bibnamefont {Roberts}},\ }\bibfield  {title}
  {\enquote {\bibinfo {title} {{Self-organization of multiple
  polariton-polariton scattering in semiconductor microcavities}},}\ }\href
  {\doibase 10.1103/PhysRevB.77.115336} {\bibfield  {journal} {\bibinfo
  {journal} {Physical Review B}\ }\textbf {\bibinfo {volume} {77}},\ \bibinfo
  {pages} {115336} (\bibinfo {year} {2008})}\BibitemShut {NoStop}%
\bibitem [{\citenamefont {Lagoudakis}\ \emph {et~al.}(2008)\citenamefont
  {Lagoudakis}, \citenamefont {Wouters}, \citenamefont {Richard}, \citenamefont
  {Baas}, \citenamefont {Carusotto}, \citenamefont {Andre}, \citenamefont
  {Dang},\ and\ \citenamefont
  {Deveaud-Pledran}}]{Lagoudakis2008QuatizedVortices}%
  \BibitemOpen
  \bibfield  {author} {\bibinfo {author} {\bibfnamefont {K.~G.}\ \bibnamefont
  {Lagoudakis}}, \bibinfo {author} {\bibfnamefont {M.}~\bibnamefont {Wouters}},
  \bibinfo {author} {\bibfnamefont {M.}~\bibnamefont {Richard}}, \bibinfo
  {author} {\bibfnamefont {A.}~\bibnamefont {Baas}}, \bibinfo {author}
  {\bibfnamefont {I.}~\bibnamefont {Carusotto}}, \bibinfo {author}
  {\bibfnamefont {R.}~\bibnamefont {Andre}}, \bibinfo {author} {\bibfnamefont
  {Le~Si}\ \bibnamefont {Dang}}, \ and\ \bibinfo {author} {\bibfnamefont
  {B.}~\bibnamefont {Deveaud-Pledran}},\ }\bibfield  {title} {\enquote
  {\bibinfo {title} {{Quantized vortices in an exciton-polariton
  condensate}},}\ }\href {\doibase 10.1038/nphys1051} {\bibfield  {journal}
  {\bibinfo  {journal} {Nature Physics}\ }\textbf {\bibinfo {volume} {4}},\
  \bibinfo {pages} {706--710} (\bibinfo {year} {2008})}\BibitemShut {NoStop}%
\bibitem [{\citenamefont {Tosi}\ \emph {et~al.}(2012)\citenamefont {Tosi},
  \citenamefont {Christmann}, \citenamefont {Berloff}, \citenamefont {Tsotsis},
  \citenamefont {Gao}, \citenamefont {Hatzopoulos}, \citenamefont {Savvidis},\
  and\ \citenamefont {Baumberg}}]{Tosi2012}%
  \BibitemOpen
  \bibfield  {author} {\bibinfo {author} {\bibfnamefont {G.}~\bibnamefont
  {Tosi}}, \bibinfo {author} {\bibfnamefont {G.}~\bibnamefont {Christmann}},
  \bibinfo {author} {\bibfnamefont {N.~G.}\ \bibnamefont {Berloff}}, \bibinfo
  {author} {\bibfnamefont {P.}~\bibnamefont {Tsotsis}}, \bibinfo {author}
  {\bibfnamefont {T.}~\bibnamefont {Gao}}, \bibinfo {author} {\bibfnamefont
  {Z.}~\bibnamefont {Hatzopoulos}}, \bibinfo {author} {\bibfnamefont {P.~G.}\
  \bibnamefont {Savvidis}}, \ and\ \bibinfo {author} {\bibfnamefont {J.~J.}\
  \bibnamefont {Baumberg}},\ }\bibfield  {title} {\enquote {\bibinfo {title}
  {{Geometrically locked vortex lattices in semiconductor quantum fluids}},}\
  }\href {\doibase 10.1038/ncomms2255} {\bibfield  {journal} {\bibinfo
  {journal} {Nature Communications}\ }\textbf {\bibinfo {volume} {3}},\
  \bibinfo {pages} {1243} (\bibinfo {year} {2012})}\BibitemShut {NoStop}%
\bibitem [{\citenamefont {Amo}\ \emph {et~al.}(2011)\citenamefont {Amo},
  \citenamefont {Pigeon}, \citenamefont {Sanvitto}, \citenamefont {Sala},
  \citenamefont {Hivet}, \citenamefont {Carusotto}, \citenamefont {Pisanello},
  \citenamefont {Lem{\'{e}}nager}, \citenamefont {Houdr{\'{e}}}, \citenamefont
  {Giacobino}, \citenamefont {Ciuti},\ and\ \citenamefont {Bramati}}]{Amo2011}%
  \BibitemOpen
  \bibfield  {author} {\bibinfo {author} {\bibfnamefont {A.}~\bibnamefont
  {Amo}}, \bibinfo {author} {\bibfnamefont {S.}~\bibnamefont {Pigeon}},
  \bibinfo {author} {\bibfnamefont {D.}~\bibnamefont {Sanvitto}}, \bibinfo
  {author} {\bibfnamefont {V.~G.}\ \bibnamefont {Sala}}, \bibinfo {author}
  {\bibfnamefont {R.}~\bibnamefont {Hivet}}, \bibinfo {author} {\bibfnamefont
  {I.}~\bibnamefont {Carusotto}}, \bibinfo {author} {\bibfnamefont
  {F.}~\bibnamefont {Pisanello}}, \bibinfo {author} {\bibfnamefont
  {G.}~\bibnamefont {Lem{\'{e}}nager}}, \bibinfo {author} {\bibfnamefont
  {R.}~\bibnamefont {Houdr{\'{e}}}}, \bibinfo {author} {\bibfnamefont
  {E.}~\bibnamefont {Giacobino}}, \bibinfo {author} {\bibfnamefont
  {C.}~\bibnamefont {Ciuti}}, \ and\ \bibinfo {author} {\bibfnamefont
  {A.}~\bibnamefont {Bramati}},\ }\bibfield  {title} {\enquote {\bibinfo
  {title} {{Polariton superfluids reveal quantum hydrodynamic solitons}},}\
  }\href {\doibase 10.1126/science.1202307} {\bibfield  {journal} {\bibinfo
  {journal} {Science}\ }\textbf {\bibinfo {volume} {332}},\ \bibinfo {pages}
  {1167--70} (\bibinfo {year} {2011})}\BibitemShut {NoStop}%
\bibitem [{\citenamefont {Sich}\ \emph {et~al.}(2012)\citenamefont {Sich},
  \citenamefont {Krizhanovskii}, \citenamefont {Skolnick}, \citenamefont
  {Gorbach}, \citenamefont {Hartley}, \citenamefont {Skryabin}, \citenamefont
  {Cerda-M{\'{e}}ndez}, \citenamefont {Biermann}, \citenamefont {Hey},\ and\
  \citenamefont {Santos}}]{Sich2011}%
  \BibitemOpen
  \bibfield  {author} {\bibinfo {author} {\bibfnamefont {M.}~\bibnamefont
  {Sich}}, \bibinfo {author} {\bibfnamefont {D.~N.}\ \bibnamefont
  {Krizhanovskii}}, \bibinfo {author} {\bibfnamefont {M.~S.}\ \bibnamefont
  {Skolnick}}, \bibinfo {author} {\bibfnamefont {A.~V.}\ \bibnamefont
  {Gorbach}}, \bibinfo {author} {\bibfnamefont {R.}~\bibnamefont {Hartley}},
  \bibinfo {author} {\bibfnamefont {D.~V.}\ \bibnamefont {Skryabin}}, \bibinfo
  {author} {\bibfnamefont {E.~A.}\ \bibnamefont {Cerda-M{\'{e}}ndez}}, \bibinfo
  {author} {\bibfnamefont {K.}~\bibnamefont {Biermann}}, \bibinfo {author}
  {\bibfnamefont {R.}~\bibnamefont {Hey}}, \ and\ \bibinfo {author}
  {\bibfnamefont {P.~V.}\ \bibnamefont {Santos}},\ }\bibfield  {title}
  {\enquote {\bibinfo {title} {{Observation of bright polariton solitons in a
  semiconductor microcavity}},}\ }\href {\doibase 10.1038/nphoton.2011.267}
  {\bibfield  {journal} {\bibinfo  {journal} {Nature Photonics}\ }\textbf
  {\bibinfo {volume} {6}},\ \bibinfo {pages} {50--55} (\bibinfo {year}
  {2012})}\BibitemShut {NoStop}%
\bibitem [{\citenamefont {Chana}\ \emph {et~al.}(2015)\citenamefont {Chana},
  \citenamefont {Sich}, \citenamefont {Fras}, \citenamefont {Gorbach},
  \citenamefont {Skryabin}, \citenamefont {Cancellieri}, \citenamefont
  {Cerda-M{\'{e}}ndez}, \citenamefont {Biermann}, \citenamefont {Hey},
  \citenamefont {Santos}, \citenamefont {Skolnick},\ and\ \citenamefont
  {Krizhanovskii}}]{Chana2015}%
  \BibitemOpen
  \bibfield  {author} {\bibinfo {author} {\bibfnamefont {J.~K.}\ \bibnamefont
  {Chana}}, \bibinfo {author} {\bibfnamefont {M.}~\bibnamefont {Sich}},
  \bibinfo {author} {\bibfnamefont {F.}~\bibnamefont {Fras}}, \bibinfo {author}
  {\bibfnamefont {A.~V.}\ \bibnamefont {Gorbach}}, \bibinfo {author}
  {\bibfnamefont {D.~V.}\ \bibnamefont {Skryabin}}, \bibinfo {author}
  {\bibfnamefont {E.}~\bibnamefont {Cancellieri}}, \bibinfo {author}
  {\bibfnamefont {E.~A.}\ \bibnamefont {Cerda-M{\'{e}}ndez}}, \bibinfo {author}
  {\bibfnamefont {K.}~\bibnamefont {Biermann}}, \bibinfo {author}
  {\bibfnamefont {R.}~\bibnamefont {Hey}}, \bibinfo {author} {\bibfnamefont
  {P.~V.}\ \bibnamefont {Santos}}, \bibinfo {author} {\bibfnamefont {M.~S.}\
  \bibnamefont {Skolnick}}, \ and\ \bibinfo {author} {\bibfnamefont {D.~N.}\
  \bibnamefont {Krizhanovskii}},\ }\bibfield  {title} {\enquote {\bibinfo
  {title} {{Spatial patterns of dissipative polariton solitons in semiconductor
  microcavities}},}\ }\href {\doibase 10.1103/PhysRevLett.115.256401}
  {\bibfield  {journal} {\bibinfo  {journal} {Physical Review Letters}\
  }\textbf {\bibinfo {volume} {115}},\ \bibinfo {pages} {256401} (\bibinfo
  {year} {2015})}\BibitemShut {NoStop}%
\bibitem [{\citenamefont {Stevenson}\ \emph {et~al.}(2000)\citenamefont
  {Stevenson}, \citenamefont {Astratov}, \citenamefont {Skolnick},
  \citenamefont {Whittaker}, \citenamefont {Emam-Ismail}, \citenamefont
  {Tartakovskii}, \citenamefont {Savvidis}, \citenamefont {Baumberg},\ and\
  \citenamefont {Roberts}}]{Stevenson2000}%
  \BibitemOpen
  \bibfield  {author} {\bibinfo {author} {\bibfnamefont {R.~M.}\ \bibnamefont
  {Stevenson}}, \bibinfo {author} {\bibfnamefont {V.~N.}\ \bibnamefont
  {Astratov}}, \bibinfo {author} {\bibfnamefont {M.~S.}\ \bibnamefont
  {Skolnick}}, \bibinfo {author} {\bibfnamefont {D.~M.}\ \bibnamefont
  {Whittaker}}, \bibinfo {author} {\bibfnamefont {M.}~\bibnamefont
  {Emam-Ismail}}, \bibinfo {author} {\bibfnamefont {A.~I.}\ \bibnamefont
  {Tartakovskii}}, \bibinfo {author} {\bibfnamefont {P.~G.}\ \bibnamefont
  {Savvidis}}, \bibinfo {author} {\bibfnamefont {J.~J.}\ \bibnamefont
  {Baumberg}}, \ and\ \bibinfo {author} {\bibfnamefont {J.~S.}\ \bibnamefont
  {Roberts}},\ }\bibfield  {title} {\enquote {\bibinfo {title} {{Continuous
  wave observation of massive polariton redistribution by stimulated scattering
  in semiconductor microcavities}},}\ }\href {\doibase
  10.1103/PhysRevLett.85.3680} {\bibfield  {journal} {\bibinfo  {journal}
  {Physical Review Letters}\ }\textbf {\bibinfo {volume} {85}},\ \bibinfo
  {pages} {3680--3683} (\bibinfo {year} {2000})}\BibitemShut {NoStop}%
\bibitem [{\citenamefont {Ferrier}\ \emph {et~al.}(2011)\citenamefont
  {Ferrier}, \citenamefont {Wertz}, \citenamefont {Johne}, \citenamefont
  {Solnyshkov}, \citenamefont {Senellart}, \citenamefont {Sagnes},
  \citenamefont {Lema{\^{i}}tre}, \citenamefont {Malpuech},\ and\ \citenamefont
  {Bloch}}]{Ferrier2011}%
  \BibitemOpen
  \bibfield  {author} {\bibinfo {author} {\bibfnamefont {L.}~\bibnamefont
  {Ferrier}}, \bibinfo {author} {\bibfnamefont {E.}~\bibnamefont {Wertz}},
  \bibinfo {author} {\bibfnamefont {R.}~\bibnamefont {Johne}}, \bibinfo
  {author} {\bibfnamefont {D.~D.}\ \bibnamefont {Solnyshkov}}, \bibinfo
  {author} {\bibfnamefont {P.}~\bibnamefont {Senellart}}, \bibinfo {author}
  {\bibfnamefont {I.}~\bibnamefont {Sagnes}}, \bibinfo {author} {\bibfnamefont
  {A.}~\bibnamefont {Lema{\^{i}}tre}}, \bibinfo {author} {\bibfnamefont
  {G.}~\bibnamefont {Malpuech}}, \ and\ \bibinfo {author} {\bibfnamefont
  {J.}~\bibnamefont {Bloch}},\ }\bibfield  {title} {\enquote {\bibinfo {title}
  {{Interactions in confined polariton condensates}},}\ }\href {\doibase
  10.1103/PhysRevLett.106.126401} {\bibfield  {journal} {\bibinfo  {journal}
  {Physical Review Letters}\ }\textbf {\bibinfo {volume} {106}},\ \bibinfo
  {pages} {126401} (\bibinfo {year} {2011})}\BibitemShut {NoStop}%
\bibitem [{\citenamefont {Wertz}\ \emph {et~al.}(2010)\citenamefont {Wertz},
  \citenamefont {Ferrier}, \citenamefont {Solnyshkov}, \citenamefont {Johne},
  \citenamefont {Sanvitto}, \citenamefont {Lema{\^{i}}tre}, \citenamefont
  {Sagnes}, \citenamefont {Grousson}, \citenamefont {Kavokin}, \citenamefont
  {Senellart}, \citenamefont {Malpuech},\ and\ \citenamefont
  {Bloch}}]{Wertz2010}%
  \BibitemOpen
  \bibfield  {author} {\bibinfo {author} {\bibfnamefont {E.}~\bibnamefont
  {Wertz}}, \bibinfo {author} {\bibfnamefont {L.}~\bibnamefont {Ferrier}},
  \bibinfo {author} {\bibfnamefont {D.~D.}\ \bibnamefont {Solnyshkov}},
  \bibinfo {author} {\bibfnamefont {R.}~\bibnamefont {Johne}}, \bibinfo
  {author} {\bibfnamefont {D.}~\bibnamefont {Sanvitto}}, \bibinfo {author}
  {\bibfnamefont {A.}~\bibnamefont {Lema{\^{i}}tre}}, \bibinfo {author}
  {\bibfnamefont {I.}~\bibnamefont {Sagnes}}, \bibinfo {author} {\bibfnamefont
  {R.}~\bibnamefont {Grousson}}, \bibinfo {author} {\bibfnamefont {A.~V.}\
  \bibnamefont {Kavokin}}, \bibinfo {author} {\bibfnamefont {P.}~\bibnamefont
  {Senellart}}, \bibinfo {author} {\bibfnamefont {G.}~\bibnamefont {Malpuech}},
  \ and\ \bibinfo {author} {\bibfnamefont {J.}~\bibnamefont {Bloch}},\
  }\bibfield  {title} {\enquote {\bibinfo {title} {{Spontaneous formation and
  optical manipulation of extended polariton condensates}},}\ }\href {\doibase
  10.1038/nphys1750} {\bibfield  {journal} {\bibinfo  {journal} {Nature
  Physics}\ }\textbf {\bibinfo {volume} {6}},\ \bibinfo {pages} {860--864}
  (\bibinfo {year} {2010})}\BibitemShut {NoStop}%
\bibitem [{\citenamefont {Dasbach}\ \emph {et~al.}(2002)\citenamefont
  {Dasbach}, \citenamefont {Schwab}, \citenamefont {Bayer}, \citenamefont
  {Krizhanovskii},\ and\ \citenamefont {Forchel}}]{Dasbach2002}%
  \BibitemOpen
  \bibfield  {author} {\bibinfo {author} {\bibfnamefont {G.}~\bibnamefont
  {Dasbach}}, \bibinfo {author} {\bibfnamefont {M.}~\bibnamefont {Schwab}},
  \bibinfo {author} {\bibfnamefont {M.}~\bibnamefont {Bayer}}, \bibinfo
  {author} {\bibfnamefont {D.~N.}\ \bibnamefont {Krizhanovskii}}, \ and\
  \bibinfo {author} {\bibfnamefont {A.}~\bibnamefont {Forchel}},\ }\bibfield
  {title} {\enquote {\bibinfo {title} {{Tailoring the polariton dispersion by
  optical confinement: Access to a manifold of elastic polariton pair
  scattering channels}},}\ }\href {\doibase 10.1103/PhysRevB.66.201201}
  {\bibfield  {journal} {\bibinfo  {journal} {Physical Review B}\ }\textbf
  {\bibinfo {volume} {66}},\ \bibinfo {pages} {201201} (\bibinfo {year}
  {2002})}\BibitemShut {NoStop}%
\bibitem [{\citenamefont {Sigurdsson}\ \emph {et~al.}(2015)\citenamefont
  {Sigurdsson}, \citenamefont {Shelykh},\ and\ \citenamefont
  {Liew}}]{Sigurdsson2015SwitchingCondensates}%
  \BibitemOpen
  \bibfield  {author} {\bibinfo {author} {\bibfnamefont {H.}~\bibnamefont
  {Sigurdsson}}, \bibinfo {author} {\bibfnamefont {I.~A.}\ \bibnamefont
  {Shelykh}}, \ and\ \bibinfo {author} {\bibfnamefont {T.~C.~H.}\ \bibnamefont
  {Liew}},\ }\bibfield  {title} {\enquote {\bibinfo {title} {{Switching waves
  in multilevel incoherently driven polariton condensates}},}\ }\href {\doibase
  10.1103/PhysRevB.92.195409} {\bibfield  {journal} {\bibinfo  {journal}
  {Physical Review B}\ }\textbf {\bibinfo {volume} {92}},\ \bibinfo {pages}
  {195409} (\bibinfo {year} {2015})}\BibitemShut {NoStop}%
\bibitem [{\citenamefont {Sigurdsson}\ \emph {et~al.}(2017)\citenamefont
  {Sigurdsson}, \citenamefont {Liew},\ and\ \citenamefont
  {Shelykh}}]{Sigurdsson2017ParityChannels}%
  \BibitemOpen
  \bibfield  {author} {\bibinfo {author} {\bibfnamefont {H.}~\bibnamefont
  {Sigurdsson}}, \bibinfo {author} {\bibfnamefont {T.~C.~H.}\ \bibnamefont
  {Liew}}, \ and\ \bibinfo {author} {\bibfnamefont {I.~A.}\ \bibnamefont
  {Shelykh}},\ }\bibfield  {title} {\enquote {\bibinfo {title} {{Parity
  solitons in nonresonantly driven-dissipative condensate channels}},}\ }\href
  {\doibase 10.1103/PhysRevB.96.205406} {\bibfield  {journal} {\bibinfo
  {journal} {Physical Review B}\ }\textbf {\bibinfo {volume} {96}},\ \bibinfo
  {pages} {205406} (\bibinfo {year} {2017})}\BibitemShut {NoStop}%
\bibitem [{\citenamefont {Tartakovskii}\ \emph {et~al.}(2000)\citenamefont
  {Tartakovskii}, \citenamefont {Krizhanovskii},\ and\ \citenamefont
  {Kulakovskii}}]{Tartakovskii2000Polariton-polaritonCase}%
  \BibitemOpen
  \bibfield  {author} {\bibinfo {author} {\bibfnamefont {A.~I.}\ \bibnamefont
  {Tartakovskii}}, \bibinfo {author} {\bibfnamefont {D.~N.}\ \bibnamefont
  {Krizhanovskii}}, \ and\ \bibinfo {author} {\bibfnamefont {V.~D.}\
  \bibnamefont {Kulakovskii}},\ }\bibfield  {title} {\enquote {\bibinfo {title}
  {{Polariton-polariton scattering in semiconductor microcavities: Distinctive
  features and similarities to the three-dimensional case}},}\ }\href {\doibase
  10.1103/PhysRevB.62.R13298} {\bibfield  {journal} {\bibinfo  {journal}
  {Physical Review B}\ }\textbf {\bibinfo {volume} {62}},\ \bibinfo {pages}
  {R13298--R13301} (\bibinfo {year} {2000})}\BibitemShut {NoStop}%
\bibitem [{SM()}]{SM}%
  \BibitemOpen
  \href@noop {} {\enquote {\bibinfo {title} {{See Supplemental Material at
  [link] for details of numerical modelling of kinetic evolution of polariton
  wavepackets.}}}\ }\BibitemShut {NoStop}%
\bibitem [{\citenamefont {Skryabin}\ \emph {et~al.}(2017)\citenamefont
  {Skryabin}, \citenamefont {Kartashov}, \citenamefont {Egorov}, \citenamefont
  {Sich}, \citenamefont {Chana}, \citenamefont {Tapia~Rodriguez}, \citenamefont
  {Walker}, \citenamefont {Clarke}, \citenamefont {Royall}, \citenamefont
  {Skolnick},\ and\ \citenamefont {Krizhanovskii}}]{Skryabin2017BackwardWire}%
  \BibitemOpen
  \bibfield  {author} {\bibinfo {author} {\bibfnamefont {D.~V.}\ \bibnamefont
  {Skryabin}}, \bibinfo {author} {\bibfnamefont {Y.~V.}\ \bibnamefont
  {Kartashov}}, \bibinfo {author} {\bibfnamefont {O.~A.}\ \bibnamefont
  {Egorov}}, \bibinfo {author} {\bibfnamefont {M.}~\bibnamefont {Sich}},
  \bibinfo {author} {\bibfnamefont {J.~K.}\ \bibnamefont {Chana}}, \bibinfo
  {author} {\bibfnamefont {L.~E.}\ \bibnamefont {Tapia~Rodriguez}}, \bibinfo
  {author} {\bibfnamefont {P.~M.}\ \bibnamefont {Walker}}, \bibinfo {author}
  {\bibfnamefont {E.}~\bibnamefont {Clarke}}, \bibinfo {author} {\bibfnamefont
  {B.}~\bibnamefont {Royall}}, \bibinfo {author} {\bibfnamefont {M.~S.}\
  \bibnamefont {Skolnick}}, \ and\ \bibinfo {author} {\bibfnamefont {D.~N.}\
  \bibnamefont {Krizhanovskii}},\ }\bibfield  {title} {\enquote {\bibinfo
  {title} {{Backward Cherenkov radiation emitted by polariton solitons in a
  microcavity wire}},}\ }\href {\doibase 10.1038/s41467-017-01751-6} {\bibfield
   {journal} {\bibinfo  {journal} {Nature Communications}\ }\textbf {\bibinfo
  {volume} {8}},\ \bibinfo {pages} {1554} (\bibinfo {year} {2017})}\BibitemShut
  {NoStop}%
\bibitem [{\citenamefont {Tinkler}\ \emph {et~al.}(2015)\citenamefont
  {Tinkler}, \citenamefont {Walker}, \citenamefont {Clarke}, \citenamefont
  {Krizhanovskii}, \citenamefont {Bastiman}, \citenamefont {Durska},\ and\
  \citenamefont {Skolnick}}]{Tinkler2015}%
  \BibitemOpen
  \bibfield  {author} {\bibinfo {author} {\bibfnamefont {L.}~\bibnamefont
  {Tinkler}}, \bibinfo {author} {\bibfnamefont {P.~M.}\ \bibnamefont {Walker}},
  \bibinfo {author} {\bibfnamefont {E.}~\bibnamefont {Clarke}}, \bibinfo
  {author} {\bibfnamefont {D.~N.}\ \bibnamefont {Krizhanovskii}}, \bibinfo
  {author} {\bibfnamefont {F.}~\bibnamefont {Bastiman}}, \bibinfo {author}
  {\bibfnamefont {M.}~\bibnamefont {Durska}}, \ and\ \bibinfo {author}
  {\bibfnamefont {M.~S.}\ \bibnamefont {Skolnick}},\ }\bibfield  {title}
  {\enquote {\bibinfo {title} {{Design and characterization of high optical
  quality InGaAs/GaAs/AlGaAs-based polariton microcavities}},}\ }\href
  {\doibase 10.1063/1.4905907} {\bibfield  {journal} {\bibinfo  {journal}
  {Applied Physics Letters}\ }\textbf {\bibinfo {volume} {106}},\ \bibinfo
  {pages} {021109} (\bibinfo {year} {2015})}\BibitemShut {NoStop}%
\bibitem [{\citenamefont {Ant{\'{o}}n}\ \emph {et~al.}(2013)\citenamefont
  {Ant{\'{o}}n}, \citenamefont {Liew}, \citenamefont {Tosi}, \citenamefont
  {Mart{\'{i}}n}, \citenamefont {Gao}, \citenamefont {Hatzopoulos},
  \citenamefont {Eldridge}, \citenamefont {Savvidis},\ and\ \citenamefont
  {Vi{\~{n}}a}}]{Anton2013}%
  \BibitemOpen
  \bibfield  {author} {\bibinfo {author} {\bibfnamefont {C.}~\bibnamefont
  {Ant{\'{o}}n}}, \bibinfo {author} {\bibfnamefont {T.~C.~H.}\ \bibnamefont
  {Liew}}, \bibinfo {author} {\bibfnamefont {G.}~\bibnamefont {Tosi}}, \bibinfo
  {author} {\bibfnamefont {M.~D.}\ \bibnamefont {Mart{\'{i}}n}}, \bibinfo
  {author} {\bibfnamefont {T.}~\bibnamefont {Gao}}, \bibinfo {author}
  {\bibfnamefont {Z.}~\bibnamefont {Hatzopoulos}}, \bibinfo {author}
  {\bibfnamefont {P.~S.}\ \bibnamefont {Eldridge}}, \bibinfo {author}
  {\bibfnamefont {P.~G.}\ \bibnamefont {Savvidis}}, \ and\ \bibinfo {author}
  {\bibfnamefont {L.}~\bibnamefont {Vi{\~{n}}a}},\ }\bibfield  {title}
  {\enquote {\bibinfo {title} {{Energy relaxation of exciton-polariton
  condensates in quasi-one-dimensional microcavities}},}\ }\href {\doibase
  10.1103/PhysRevB.88.035313} {\bibfield  {journal} {\bibinfo  {journal}
  {Physical Review B}\ }\textbf {\bibinfo {volume} {88}},\ \bibinfo {pages}
  {035313} (\bibinfo {year} {2013})}\BibitemShut {NoStop}%
\bibitem [{\citenamefont {Hammani}\ \emph {et~al.}(2010)\citenamefont
  {Hammani}, \citenamefont {Kibler}, \citenamefont {Finot},\ and\ \citenamefont
  {Picozzi}}]{Hammani2010}%
  \BibitemOpen
  \bibfield  {author} {\bibinfo {author} {\bibfnamefont {K.}~\bibnamefont
  {Hammani}}, \bibinfo {author} {\bibfnamefont {B.}~\bibnamefont {Kibler}},
  \bibinfo {author} {\bibfnamefont {C.}~\bibnamefont {Finot}}, \ and\ \bibinfo
  {author} {\bibfnamefont {A.}~\bibnamefont {Picozzi}},\ }\bibfield  {title}
  {\enquote {\bibinfo {title} {{Emergence of rogue waves from optical
  turbulence}},}\ }\href {\doibase 10.1016/j.physleta.2010.06.035} {\bibfield
  {journal} {\bibinfo  {journal} {Physics Letters A}\ }\textbf {\bibinfo
  {volume} {374}},\ \bibinfo {pages} {3585--3589} (\bibinfo {year}
  {2010})}\BibitemShut {NoStop}%
\bibitem [{\citenamefont {Skryabin}\ \emph {et~al.}(2003)\citenamefont
  {Skryabin}, \citenamefont {Luan}, \citenamefont {Knight},\ and\ \citenamefont
  {Russell}}]{Skryabin2003}%
  \BibitemOpen
  \bibfield  {author} {\bibinfo {author} {\bibfnamefont {D.~V.}\ \bibnamefont
  {Skryabin}}, \bibinfo {author} {\bibfnamefont {F.}~\bibnamefont {Luan}},
  \bibinfo {author} {\bibfnamefont {J.~C.}\ \bibnamefont {Knight}}, \ and\
  \bibinfo {author} {\bibfnamefont {P.~St.~J.}\ \bibnamefont {Russell}},\
  }\bibfield  {title} {\enquote {\bibinfo {title} {{Soliton self-frequency
  shift cancellation in photonic crystal fibers}},}\ }\href
  {http://science.sciencemag.org/content/301/5640/1705/tab-article-info}
  {\bibfield  {journal} {\bibinfo  {journal} {Science}\ }\textbf {\bibinfo
  {volume} {301}},\ \bibinfo {pages} {1705--1708} (\bibinfo {year}
  {2003})}\BibitemShut {NoStop}%
\bibitem [{\citenamefont {Walker}\ \emph {et~al.}(2017)\citenamefont {Walker},
  \citenamefont {Tinkler}, \citenamefont {Royall}, \citenamefont {Skryabin},
  \citenamefont {Farrer}, \citenamefont {Ritchie}, \citenamefont {Skolnick},\
  and\ \citenamefont {Krizhanovskii}}]{Walker2017}%
  \BibitemOpen
  \bibfield  {author} {\bibinfo {author} {\bibfnamefont {P.~M.}\ \bibnamefont
  {Walker}}, \bibinfo {author} {\bibfnamefont {L.}~\bibnamefont {Tinkler}},
  \bibinfo {author} {\bibfnamefont {B.}~\bibnamefont {Royall}}, \bibinfo
  {author} {\bibfnamefont {D.~V.}\ \bibnamefont {Skryabin}}, \bibinfo {author}
  {\bibfnamefont {I.}~\bibnamefont {Farrer}}, \bibinfo {author} {\bibfnamefont
  {D.~A.}\ \bibnamefont {Ritchie}}, \bibinfo {author} {\bibfnamefont {M.~S.}\
  \bibnamefont {Skolnick}}, \ and\ \bibinfo {author} {\bibfnamefont {D.~N.}\
  \bibnamefont {Krizhanovskii}},\ }\bibfield  {title} {\enquote {\bibinfo
  {title} {{Dark solitons in high velocity waveguide polariton fluids}},}\
  }\href {\doibase 10.1103/PhysRevLett.119.097403} {\bibfield  {journal}
  {\bibinfo  {journal} {Physical Review Letters}\ }\textbf {\bibinfo {volume}
  {119}},\ \bibinfo {pages} {097403} (\bibinfo {year} {2017})}\BibitemShut
  {NoStop}%
\bibitem [{\citenamefont {Krizhanovskii}\ \emph {et~al.}(2001)\citenamefont
  {Krizhanovskii}, \citenamefont {Tartakovskii}, \citenamefont {Chernenko},
  \citenamefont {Kulakovskii}, \citenamefont {Emam-Ismail}, \citenamefont
  {Skolnick},\ and\ \citenamefont
  {Roberts}}]{Krizhanovskii2001EnergyMicrocavities}%
  \BibitemOpen
  \bibfield  {author} {\bibinfo {author} {\bibfnamefont {D.~N.}\ \bibnamefont
  {Krizhanovskii}}, \bibinfo {author} {\bibfnamefont {A.~I.}\ \bibnamefont
  {Tartakovskii}}, \bibinfo {author} {\bibfnamefont {A.~V.}\ \bibnamefont
  {Chernenko}}, \bibinfo {author} {\bibfnamefont {V.~D.}\ \bibnamefont
  {Kulakovskii}}, \bibinfo {author} {\bibfnamefont {M.}~\bibnamefont
  {Emam-Ismail}}, \bibinfo {author} {\bibfnamefont {M.~S.}\ \bibnamefont
  {Skolnick}}, \ and\ \bibinfo {author} {\bibfnamefont {J.~S.}\ \bibnamefont
  {Roberts}},\ }\bibfield  {title} {\enquote {\bibinfo {title} {{Energy
  relaxation of resonantly excited polaritons in semiconductor
  microcavities}},}\ }\href {\doibase 10.1016/S0038-1098(01)00159-4} {\bibfield
   {journal} {\bibinfo  {journal} {Solid State Communications}\ }\textbf
  {\bibinfo {volume} {118}},\ \bibinfo {pages} {583--587} (\bibinfo {year}
  {2001})}\BibitemShut {NoStop}%
\bibitem [{\citenamefont {Kulakovskii}\ \emph {et~al.}(2002)\citenamefont
  {Kulakovskii}, \citenamefont {Tartakovskii}, \citenamefont {Krizhanovskii},
  \citenamefont {Skolnick},\ and\ \citenamefont
  {Roberts}}]{Kulakovskii2002ImpactPolaritons}%
  \BibitemOpen
  \bibfield  {author} {\bibinfo {author} {\bibfnamefont {V.~D.}\ \bibnamefont
  {Kulakovskii}}, \bibinfo {author} {\bibfnamefont {A.~I.}\ \bibnamefont
  {Tartakovskii}}, \bibinfo {author} {\bibfnamefont {D.~N.}\ \bibnamefont
  {Krizhanovskii}}, \bibinfo {author} {\bibfnamefont {M.~S.}\ \bibnamefont
  {Skolnick}}, \ and\ \bibinfo {author} {\bibfnamefont {J.~S.}\ \bibnamefont
  {Roberts}},\ }\bibfield  {title} {\enquote {\bibinfo {title} {{Impact of
  exciton localization on cavity polaritons}},}\ }\href {\doibase
  10.1016/S1386-9477(02)00162-5} {\bibfield  {journal} {\bibinfo  {journal}
  {Physica E: Low-Dimensional Systems and Nanostructures}\ }\textbf {\bibinfo
  {volume} {13}},\ \bibinfo {pages} {455--458} (\bibinfo {year}
  {2002})}\BibitemShut {NoStop}%
\bibitem [{\citenamefont {Tartakovskii}\ \emph {et~al.}(2003)\citenamefont
  {Tartakovskii}, \citenamefont {Krizhanovskii}, \citenamefont {Malpuech},
  \citenamefont {Emam-Ismail}, \citenamefont {Chernenko}, \citenamefont
  {Kavokin}, \citenamefont {Kulakovskii}, \citenamefont {Skolnick},\ and\
  \citenamefont {Roberts}}]{Tartakovskii2003GiantTheory}%
  \BibitemOpen
  \bibfield  {author} {\bibinfo {author} {\bibfnamefont {A.~I.}\ \bibnamefont
  {Tartakovskii}}, \bibinfo {author} {\bibfnamefont {D.~N.}\ \bibnamefont
  {Krizhanovskii}}, \bibinfo {author} {\bibfnamefont {G.}~\bibnamefont
  {Malpuech}}, \bibinfo {author} {\bibfnamefont {M.}~\bibnamefont
  {Emam-Ismail}}, \bibinfo {author} {\bibfnamefont {A.~V.}\ \bibnamefont
  {Chernenko}}, \bibinfo {author} {\bibfnamefont {A.~V.}\ \bibnamefont
  {Kavokin}}, \bibinfo {author} {\bibfnamefont {V~D}\ \bibnamefont
  {Kulakovskii}}, \bibinfo {author} {\bibfnamefont {M~S}\ \bibnamefont
  {Skolnick}}, \ and\ \bibinfo {author} {\bibfnamefont {J~S}\ \bibnamefont
  {Roberts}},\ }\bibfield  {title} {\enquote {\bibinfo {title} {{Giant
  enhancement of polariton relaxation in semiconductor microcavities by
  polariton-free carrier interaction: Experimental evidence and theory}},}\
  }\href {\doibase 10.1103/PhysRevB.67.165302} {\bibfield  {journal} {\bibinfo
  {journal} {Physical Review B}\ }\textbf {\bibinfo {volume} {67}},\ \bibinfo
  {pages} {165302} (\bibinfo {year} {2003})}\BibitemShut {NoStop}%
\bibitem [{\citenamefont {Savvidis}\ \emph {et~al.}(2002)\citenamefont
  {Savvidis}, \citenamefont {Baumberg}, \citenamefont {Porras}, \citenamefont
  {Whittaker}, \citenamefont {Skolnick},\ and\ \citenamefont
  {Roberts}}]{Savvidis2002RingMicrocavities}%
  \BibitemOpen
  \bibfield  {author} {\bibinfo {author} {\bibfnamefont {P.~G.}\ \bibnamefont
  {Savvidis}}, \bibinfo {author} {\bibfnamefont {J.~J.}\ \bibnamefont
  {Baumberg}}, \bibinfo {author} {\bibfnamefont {D.}~\bibnamefont {Porras}},
  \bibinfo {author} {\bibfnamefont {D.~M.}\ \bibnamefont {Whittaker}}, \bibinfo
  {author} {\bibfnamefont {M.~S.}\ \bibnamefont {Skolnick}}, \ and\ \bibinfo
  {author} {\bibfnamefont {J.~S.}\ \bibnamefont {Roberts}},\ }\bibfield
  {title} {\enquote {\bibinfo {title} {{Ring emission and exciton-pair
  scattering in semiconductor microcavities}},}\ }\href {\doibase
  10.1103/PhysRevB.65.073309} {\bibfield  {journal} {\bibinfo  {journal}
  {Physical Review B}\ }\textbf {\bibinfo {volume} {65}},\ \bibinfo {pages}
  {073309} (\bibinfo {year} {2002})}\BibitemShut {NoStop}%
\bibitem [{\citenamefont {Caputo}\ \emph {et~al.}(2016)\citenamefont {Caputo},
  \citenamefont {Ballarini}, \citenamefont {Dagvadorj}, \citenamefont
  {Mu{\~{n}}oz}, \citenamefont {De~Giorgi}, \citenamefont {Dominici},
  \citenamefont {West}, \citenamefont {Pfeiffer}, \citenamefont {Gigli},
  \citenamefont {Laussy}, \citenamefont {Szyma{\'{n}}ska},\ and\ \citenamefont
  {Sanvitto}}]{Caputo2016}%
  \BibitemOpen
  \bibfield  {author} {\bibinfo {author} {\bibfnamefont {D.}~\bibnamefont
  {Caputo}}, \bibinfo {author} {\bibfnamefont {D.}~\bibnamefont {Ballarini}},
  \bibinfo {author} {\bibfnamefont {G.}~\bibnamefont {Dagvadorj}}, \bibinfo
  {author} {\bibfnamefont {C.~S.}\ \bibnamefont {Mu{\~{n}}oz}}, \bibinfo
  {author} {\bibfnamefont {M.}~\bibnamefont {De~Giorgi}}, \bibinfo {author}
  {\bibfnamefont {L.}~\bibnamefont {Dominici}}, \bibinfo {author}
  {\bibfnamefont {K.}~\bibnamefont {West}}, \bibinfo {author} {\bibfnamefont
  {L.N.}\ \bibnamefont {Pfeiffer}}, \bibinfo {author} {\bibfnamefont
  {G.}~\bibnamefont {Gigli}}, \bibinfo {author} {\bibfnamefont {F.~P.}\
  \bibnamefont {Laussy}}, \bibinfo {author} {\bibfnamefont {M.~H.}\
  \bibnamefont {Szyma{\'{n}}ska}}, \ and\ \bibinfo {author} {\bibfnamefont
  {D.}~\bibnamefont {Sanvitto}},\ }\bibfield  {title} {\enquote {\bibinfo
  {title} {{Topological order and equilibrium in a condensate of
  exciton-polaritons}},}\ }\href {http://arxiv.org/abs/1610.05737} {\bibfield
  {journal} {\bibinfo  {journal} {arXiv:1610.05737}\ } (\bibinfo {year}
  {2016})}\BibitemShut {NoStop}%
\bibitem [{\citenamefont {Vladimirova}\ \emph {et~al.}(2010)\citenamefont
  {Vladimirova}, \citenamefont {Cronenberger}, \citenamefont {Scalbert},
  \citenamefont {Kavokin}, \citenamefont {Miard}, \citenamefont
  {Lema{\^{i}}tre}, \citenamefont {Bloch}, \citenamefont {Solnyshkov},
  \citenamefont {Malpuech},\ and\ \citenamefont {Kavokin}}]{Vladimirova2010}%
  \BibitemOpen
  \bibfield  {author} {\bibinfo {author} {\bibfnamefont {M.}~\bibnamefont
  {Vladimirova}}, \bibinfo {author} {\bibfnamefont {S.}~\bibnamefont
  {Cronenberger}}, \bibinfo {author} {\bibfnamefont {D.}~\bibnamefont
  {Scalbert}}, \bibinfo {author} {\bibfnamefont {K.~V.}\ \bibnamefont
  {Kavokin}}, \bibinfo {author} {\bibfnamefont {A.}~\bibnamefont {Miard}},
  \bibinfo {author} {\bibfnamefont {A.}~\bibnamefont {Lema{\^{i}}tre}},
  \bibinfo {author} {\bibfnamefont {J.}~\bibnamefont {Bloch}}, \bibinfo
  {author} {\bibfnamefont {D.}~\bibnamefont {Solnyshkov}}, \bibinfo {author}
  {\bibfnamefont {G.}~\bibnamefont {Malpuech}}, \ and\ \bibinfo {author}
  {\bibfnamefont {A.~V.}\ \bibnamefont {Kavokin}},\ }\bibfield  {title}
  {\enquote {\bibinfo {title} {{Polariton-polariton interaction constants in
  microcavities}},}\ }\href {\doibase 10.1103/PhysRevB.82.075301} {\bibfield
  {journal} {\bibinfo  {journal} {Physical Review B}\ }\textbf {\bibinfo
  {volume} {82}},\ \bibinfo {pages} {075301} (\bibinfo {year}
  {2010})}\BibitemShut {NoStop}%
\end{thebibliography}%

\end{document}